# Quantum Biosensors on Chip: A Review from Electronic and Photonic Integrated Circuits to Future Integrated Quantum Photonic Circuits

Yasaman Torabi [1,*], Shahram Shirani [1,2], James P. Reilly [1]

[1] Electrical and Computer Engineering Department, McMaster University, Hamilton, Ontario, Canada
[2] L.R. Wilson/Bell Canada Chair in Data Communications, Hamilton, Ontario, Canada

* Correspondence: torabiy@mcmaster.ca

**Abstract:** Quantum biosensors offer a promising route to overcome the sensitivity and specificity limitations of conventional biosensing technologies. Their ability to detect biochemical signals at extremely low concentrations makes them strong candidates for next-generation sensing systems. This paper reviews the current state of quantum biosensors and discusses their future implementation in chip-scale platforms that combine microelectronic and photonic technologies. It covers key quantum biosensing approaches including quantum dots, and nitrogen-vacancy (NV) centers. This paper also considers their potential compatibility with electronic integrated circuits (EICs), photonic integrated circuits (PICs) and integrated quantum photonic (IQP) systems for future biosensing applications. To our knowledge, this is the first review to systematically connect quantum biosensing technologies with the development of microelectronic and photonic chip-based devices. The goal is to clarify the technological trajectory toward compact, scalable, and high-performance quantum biosensing systems.

## 1. Introduction

Biosensing technologies have transformed clinical diagnostics, environmental monitoring, and public health surveillance. They are critical in clinical diagnostics for rapid and accurate detection of diseases, biomarkers, and pathogens, as well as in environmental monitoring for identifying pollutants, toxins, and pathogens [1]. In 1962, Clark and Lyons invented the first biosensor for the detection of blood oxygen content using enzymes. Since then, several efforts have been made to improve biosensing mechanism [2]. Advances in microfabrication and material science have facilitated the shift from traditional laboratory-scale devices to integrated chip-based sensors capable of rapid, label-free, and real-time detection. Electronic Integrated Circuits (EICs) represented the first significant breakthrough in chip-based biosensing. Since the introduction of complementary metal–oxide–semiconductor (CMOS) technology in the late 20th century, numerous EIC biosensors have been developed [3]. These devices detect biological signals by converting biochemical interactions into measurable electrical signals. EIC biosensors have demonstrated significant success, with applications including cardiovascular monitoring [4], cancer biomarker detection [5], and glucose sensing [6]. Nevertheless, fundamental limitations such as susceptibility to electrical noise, and power dissipation constraints have driven exploration into alternative integrated sensing paradigms.

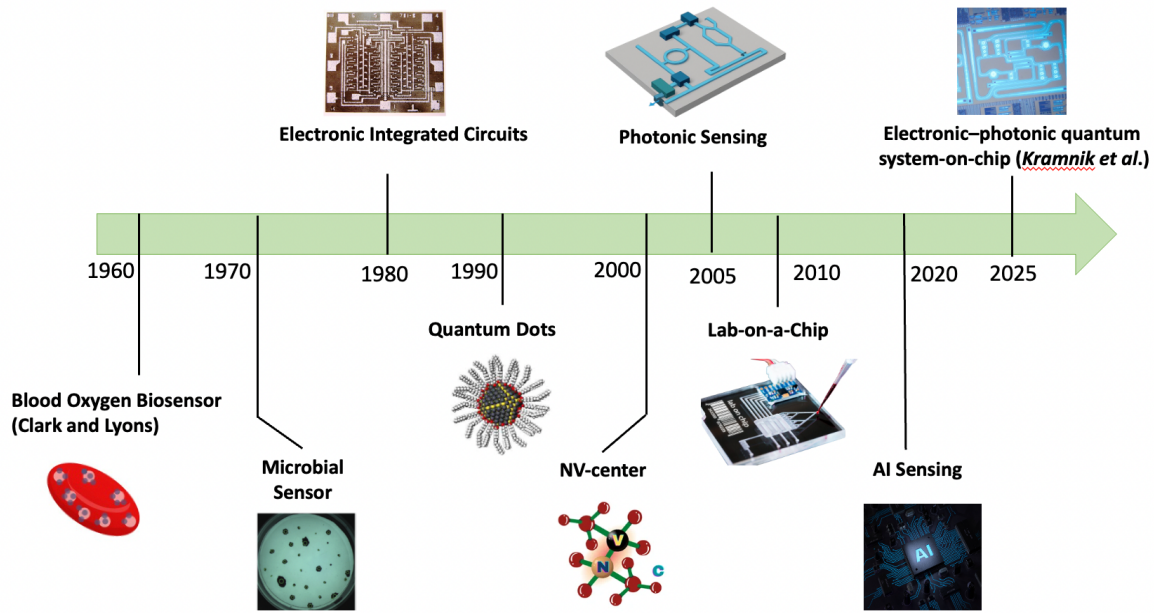

**Figure 1.** Biosensor development timeline.

Photonic Integrated Circuits (PICs) are one of the promising successors due to their advantages in speed, sensitivity, and immunity to electromagnetic interference. The first successful deployment of a commercial PIC was made in 2004 [7]. Since their inception in the early 2000s, PIC biosensors have utilized optical waveguides, ring resonators, and Mach–Zehnder interferometers (MZIs) to achieve high-performance biosensing capabilities. PIC-based sensors have been successfully applied in various critical domains, such as detecting cancer biomarkers [8], virus detection [9], and monitoring of water pollutants [10]. However, PIC technology still encounters challenges, including complex fabrication processes, precise optical alignment requirements, and optical nonlinearities that constrain broader practical adoption.

Parallel advancements in quantum technologies in early 2000s have introduced quantum sensing [11]. Thanks to quantum phenomena such as coherence, tunneling, and entanglement, these sensors can improve sensitivity and specificity limits in comparison with classical methods. Quantum biosensing approaches include quantum plasmonic sensors, quantum dots (QDs), and nitrogen-vacancy (NV) centers, each providing unique sensing mechanisms. Plasmonic quantum sensors detect molecular interactions by monitoring changes in light absorption or scattering from metal nanoparticles; quantum dot sensors exploit size-dependent fluorescence emission for detection; and NV center-based sensors use optically readable spin states in diamond for nanoscale bioimaging applications [12–14]. Nonetheless, quantum biosensors currently face significant integration challenges, including quantum state decoherence, complex readout systems, and the requirement for cryogenic temperatures in certain cases. To address these limitations, integrated quantum photonic (IQP) biosensors have been considered as a future-oriented solution. IQP devices aim to combine quantum sensing principles with microelectronic and photonic integration on silicon platforms. In July 2025, Kramnik et al. [15] demonstrated the first integrated quantum photonic chip that combines electronic, photonic, and quantum components on a single silicon platform, fabricated in a commercial foundry for the first time, which is a major step toward IQP biosensors. Figure 1 shows biosensor development timeline. This paper explores the evolution of EIC, PIC, and quantum biosensors, highlighting their historical developments, technological advantages, existing limitations, and outlining a roadmap toward IQP biosensors.

## 2. Fundamentals of Quantum Technologies

*2.1 Introduction to Quantum Concepts*

Quantum mechanics governs the behavior of particles and energy at the atomic and subatomic scale, departing fundamentally from classical physics. These principles are central to quantum technologies that are now being applied in healthcare for sensing, imaging, and computing. One key concept that stems from these principles is the qubit. A quantum bit, or qubit, is the fundamental unit of quantum information. Unlike classical bits which exist only in definite states $|0\rangle$ or $|1\rangle$, a qubit can exist in a linear combination of both states. This phenomenon is called superposition. equation (1) defines the foundational superposition state of a two-level quantum system [16].

$$|\psi\rangle = \alpha|0\rangle + \beta|1\rangle, \quad \text{with } |\alpha|^2 + |\beta|^2 = 1, \tag{1}$$

where $\alpha$ and $\beta$ are complex probability amplitudes. Superposition allows quantum systems to encode and manipulate more information than classical devices, which is essential in quantum-accelerated tasks such as protein folding [17,18] and molecular structure search [19,20]. Another cornerstone is entanglement, in which two or more particles share a quantum state that cannot be described independently. A prototypical example is the Bell state (equation 2) [21]. Such entangled states form the foundation for quantum communication protocols and distributed sensing systems, which are increasingly being studied for biomedical applications requiring precise, correlated measurements [22,23].

$$|\Phi^+\rangle = \frac{1}{\sqrt{2}}[|00\rangle + |11\rangle]. \tag{2}$$

Another fundamental idea is the Heisenberg uncertainty principle. This is mathematically expressed as equation 3 and it imposes a lower bound on the simultaneous precision with which conjugate variables (e.g., position and momentum) can be known. This is fundamental to quantum measurement theory and directly influences the noise floor of biosensors and imaging systems [24,25]. This intrinsic uncertainty enables quantum-enhanced metrology. For instance, by exploiting squeezed states, quantum sensors can detect tiny magnetic or electric field fluctuations within biological tissues [26–28].

$$\sigma_x \sigma_p \geq \frac{\hbar}{2}. \tag{3}$$

Time evolution of quantum systems is governed by the Schrödinger equation (equation 4), a differential equation that describes how quantum states evolve over time. Here, $\hat{H}$ is the Hamiltonian operator, which encodes the energy structure of the system. Solutions to this equation form the basis for simulating quantum dynamics in biochemical systems, including reaction mechanisms and metabolic pathways [29,30].

$$i\hbar \frac{\partial}{\partial t}|\psi(t)\rangle = \hat{H}|\psi(t)\rangle. \tag{4}$$

A key advantage of quantum sensors is their ability to surpass classical precision limits. The Standard Quantum Limit (SQL) represents the best possible precision achievable using classical resources. It states that the uncertainty in estimating a parameter $\theta$ scales inversely with the square root of the number of independent measurements or particles $N$ (equation 5).

$$\Delta\theta_{\text{SQL}} \sim \frac{1}{\sqrt{N}}. \tag{5}$$

When entangled resources are used, the measurement precision approaches the Heisenberg Limit (equation 6). This improved scaling is critical in the development of highly sensitive diagnostic tools for detecting weak biosignals, such as neural electromagnetic activity [31].

$$\Delta\theta_{\text{HL}} \sim \frac{1}{N}. \tag{6}$$

These quantum principles form the theoretical infrastructure upon which quantum microelectronic healthcare systems are being designed. Understanding these foundations is essential for translating quantum mechanics into clinical and biomedical technologies.

*2.2 Quantum Phenomena in Biomedical Contexts*

Quantum phenomena in the domains of physics and chemistry are now increasingly recognized for their roles in biological processes and healthcare technologies. These include tunnelling, coherence, and entanglement, which not only explain fundamental mechanisms in living systems but also enable next-generation diagnostics, sensing, and computation.

2.1.1. Quantum Tunnelling

One of the earliest and most well-established quantum effects in biology is quantum tunnelling, where subatomic particles such as electrons or protons cross energy barriers that are classically forbidden. In enzymatic catalysis, tunnelling enhances reaction rates far beyond what thermal activation can explain [32]. This enhancement can be modeled by incorporating a tunnelling correction factor ($\kappa$) into the classical Arrhenius expression for the reaction rate (equation 7).

$$k = A \exp\left(-\frac{G^{\ddagger}}{k_B T}\right)\kappa, \tag{7}$$

where $A$ is the pre-exponential factor, $G^{\ddagger}$ is the activation free energy, $k_B$ is Boltzmann's constant, $T$ is temperature, and $\kappa$ quantifies the contribution from tunnelling. Recent studies have shown that hydrogen-transfer enzymes rely heavily on tunnelling mechanisms which influences drug metabolism [33]. Quantum models have also revisited proton tunnelling in DNA [33,34], originally proposed by Löwdin [35]. This has implications for cancer development [36] and genomic stability [37]. The role of tunnelling has also extended into virology. In 2022 and 2024, researchers modeled vibration-assisted electron tunnelling as a facilitator of SARS-CoV-2 spike protein binding to bioreceptors. By incorporating quantum state diffusion and vibrational mode coupling, these studies provided mechanistic insight into how viral proteins may exploit quantum-enhanced binding kinetics [38,39].

Figure 2 depicts the principle of quantum tunneling. In classical mechanics, a particle with energy (E) less than the barrier (V) cannot cross and is reflected back. In contrast, quantum mechanics describes the particle as a wavefunction that penetrates and decays exponentially within the barrier, even when its energy is lower than the barrier height. As shown in the graph, the wavefunction extends beyond the barrier, illustrating that there is a nonzero probability of finding the particle on the other side.

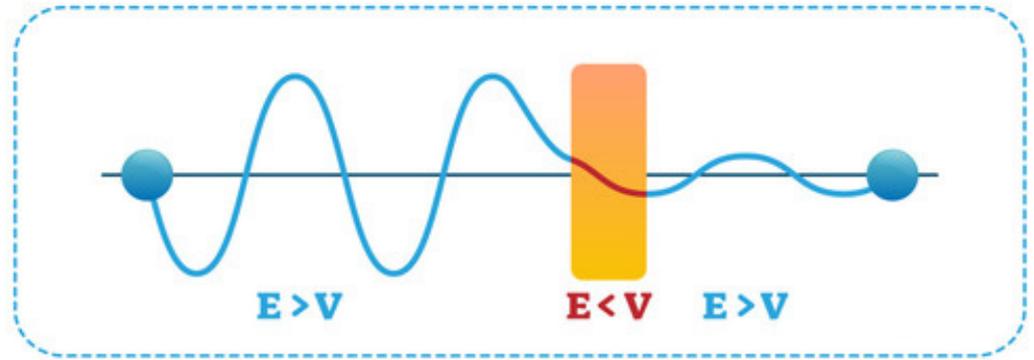

**Figure 2.** Quantum tunneling through a barrier [33].

2.1.2. Quantum Coherence

Building upon the foundational understanding of tunnelling, recent advances have turned toward quantum coherence. Quantum coherence refers to the ability of a quantum system to maintain well-defined phase relationships between its states over time. The degree of coherence in a quantum sensor is often quantified by the coherence time ($T$), which characterizes how long phase relationships are preserved (equation 8).

$$S(t) = S_0 exp\left(-\frac{t}{T}\right), \tag{8}$$

where $S(t)$ is the signal at time $t$, $S_0$ is the initial signal amplitude and T is the coherence time. This phenomenon is used in quantum sensors, particularly those based on diamond nitrogen-vacancy (NV) centers, which exhibit extreme sensitivity to magnetic and electric fields [40]. These technologies are now being explored for biomedical applications, such as wearable magnetoencephalography (MEG) [41], real-time brain activity monitoring [42], and single-cell spectroscopy [43]. For example, Figure 3 illustrates how quantum coherence may play a role in neuronal function. The diagram shows quantum interactions between protein-based qubits within the nanoscale environment of a voltage-gated potassium channel. This suggests that quantum coherence within ion channels could influence neural signaling at the cellular level [44].

2.1.3. Quantum Entanglement and Superposition

Quantum entanglement and superposition are being applied in computational healthcare. Quantum processors can model complex molecular systems [45], accelerate drug discovery pipelines [46], and optimize personalized treatment strategies [47]. A growing body of work now demonstrates the feasibility of applying quantum machine learning to tasks such as genomic analysis [48], protein folding [49], and medical image reconstruction [50].

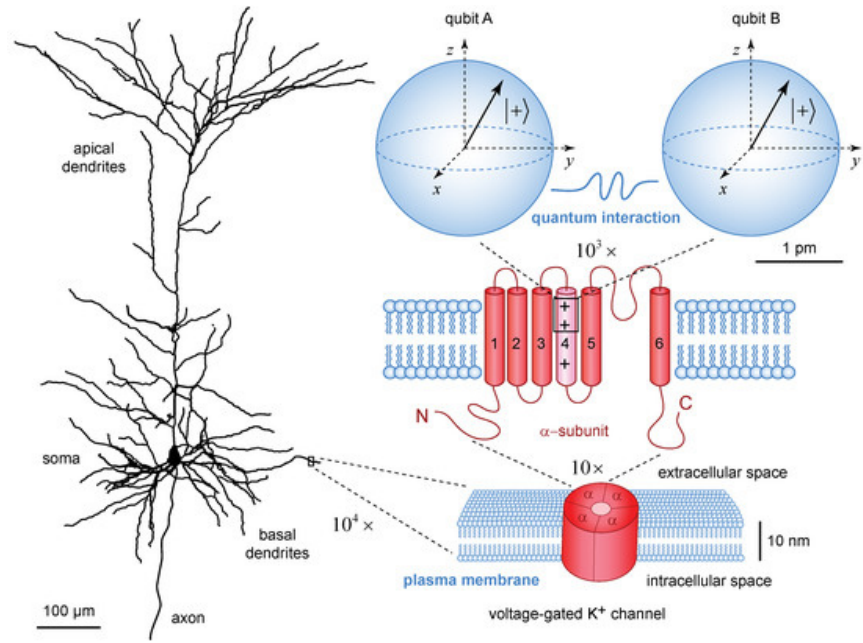

**Figure 3.** Quantum model of neural signaling [44].

*2.2 Quantum Devices: Qubits, Sensors, and Photons*

Quantum devices are central to the development of quantum technologies in both computation and sensing. The main classes including qubits, quantum sensors, and photonic systems operate based on core quantum principles. Table 1 provides a comparative overview of these quantum devices. Qubits are quantum units of information that exist in superpositions of 0 and 1. Early implementations relied on trapped ions and superconducting circuits. More recent developments include silicon spin qubits [51], diamond NV centers [52], and topological qubits [53]. Since 2016, superconducting platforms have shown considerable progress, with multi-qubit systems demonstrated by IBM and Google [54]. Solid-state qubits are also being applied in biomedical imaging and molecular sensing [55]. Quantum sensors measure physical quantities with extreme precision. For example, NV center magnetometers offer nanometer-scale resolution at room temperature [56]. Portable quantum sensors have advanced toward clinical use, such as quantum-enhanced magnetoencephalography (MEG) [57]. Photonic devices manipulate single photons for quantum information processing. Advances in integrated photonics have led to the development of on-chip entangled photon sources, and CMOS-compatible biosensors [58,59].

**Table 1.** Summary of Discussed Quantum Devices

| Class | Operating Principle | Implementations | Biomedical Applications | Advantages | Limitations |
|---|---|---|---|---|---|
| Qubits | Superposition, Entanglement | Trapped ions, Superconducting circuits, Silicon spin qubits, NV centers, Topological qubits [34-37] | Molecular sensing, Biomedical imaging [38] | Scalable, High fidelity | Decoherence, Cryogenic needs, Fabrication complexity |
| Quantum Sensors | Quantum coherence, Interference | NV center Magnetometers [56] | Brain imaging, Nano-scopic magnetic sensing [39] | High sensitivity, Room-temperature | Integration challenges, SNR issues |
| Quantum Photonic | Single-photon | On-chip photon sources, CMOS-compatible biosensors [41,42] | Optical biosensing, Quantum communication | Compact, Fast, Low noise, CMOS-compatible | Photon loss, Detection inefficiency |

*2.3 Interfacing Quantum and Classical Systems*

The operation of quantum technologies heavily depends on interfaces with classical electronics. These interfaces must reconcile differences in operating principles. Quantum systems rely on superposition, coherence, and low-temperature environments, while classical systems are deterministic, noisy, and typically ambient. Recent research has focused on developing cryogenic-compatible CMOS control electronics for superconducting and spin qubits [60–62], and hybrid optoelectronic platforms for quantum photonic systems [63,64]. Table 2 summarizes different quantum hardware factors. Among these, silicon spin qubits (Figure 4c) and photonic systems (Figure 4e) are generally the most suitable for microelectronic integration. Silicon spin qubits can be built using standard chip technology, while photonic systems work at room temperature and are already compatible with chip-based optics [64,65]. Superconducting circuits (Figure 4b) are impressive for their speed but need extremely low temperatures, which makes them harder to use in regular microelectronic setups [66]. NV centers in diamond (Figure 4d) stand out for working at room temperature, but their weak output signals limit their use [67]. Trapped ions, although very precise, still depend on large optical setups (Figure 4a) that don't translate easily to chip-scale electronics [68].

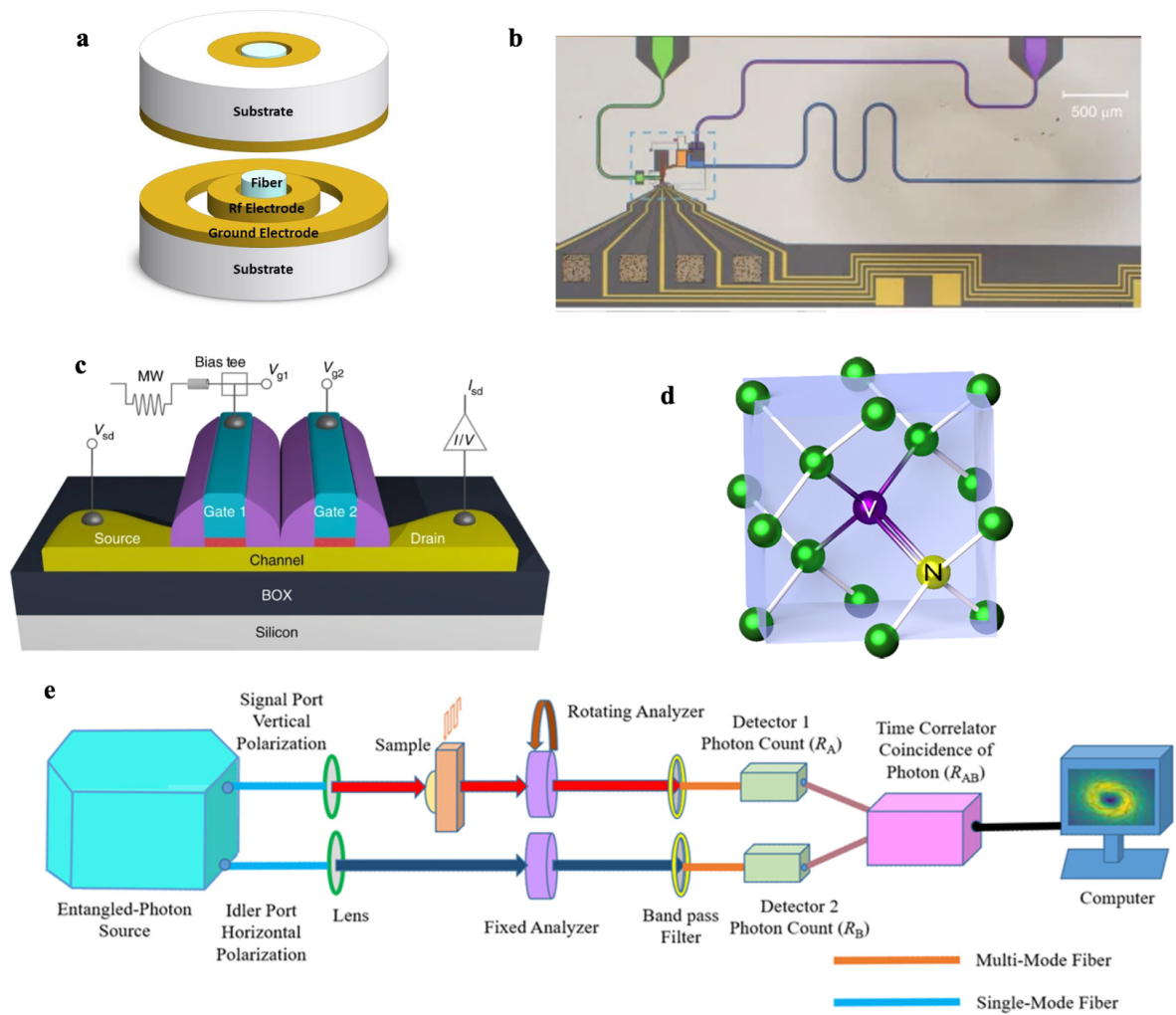

**Figure 4. (a)** Schematic of a 3D ion trap assembly [69] **(b)** Micrograph of an integrated superconducting qubit circuit [70] **(c)** Structure of a silicon spin qubit device using CMOS technology [71] **(d)** Model of a nitrogen-vacancy (NV) center defect in diamond [72] **(e)** Setup for generating and detecting entangled photon pairs [73].

Table 2. Comparison of different quantum platforms.

| Working Principle | Example | Temperature | Control Method | Latency | Pros | Cons |
|---|---|---|---|---|---|---|
| Superposition | Trapped Ions [68] | 10–100 mK | Laser pulses, RF fields | 10-500 μs | Precise control, High-fidelity entanglement | Bulky optics, Slow |
| Coherent Charge States | Superconducting Circuits [66] | 10-20 mK | Microwave pulses | 10–300 ns | Fast switching, Supports multi-qubit arrays | Low-temp, microwave crosstalk in dense arrays |
| Spin-Based | Silicon Spin Qubits [65] | 50 mK–1 K | GHz pulses | 1–10 μs | Compact, CMOS process fabrication | Low-temp, Noise-sensitive |
| Defect-Based Qubits | NV Centers [67] | 300–700 K | Optical and microwave drive | 1-500 μs | Room-temp | Weak signals, Hard on-chip integration |
| Quantum Photonic | Entangled Photons [64] | 273–300 K | Electro-optic/thermal tuning | 1–10 ns | Room-temp, Fast data rates | Complex setup, Optical loss |

## 3. Quantum Biosensing Systems

*3.1 Quantum Sensing Principles*

Quantum sensing uses quantum properties to measure physical quantities with higher sensitivity than classical sensors. A typical quantum sensing has three essential stages. First, the quantum sensor is initialized into a well-defined quantum state, ensuring consistency across trials; multiple preparations are carried out to gather statistically meaningful data. Second, the sensor coherently interacts with the target system, during which quantum features like phase shifts or spin dynamics become imprinted with information about the physical quantity of interest. Finally, the quantum state is measured repeatedly, and the collected outcomes are analyzed to extract the signal with a high signal-to-noise ratio, with precision beyond what classical systems can achieve [74,75].

In this review, we focus on three major quantum sensing approaches: plasmonic, quantum dot (QD), and nitrogen-vacancy (NV) center-based sensors. Plasmonic sensors exploit quantum tunneling and surface plasmon resonance between metallic nanoparticles[76]. Quantum dot sensors use size-tunable semiconductor nanocrystals that exhibit discrete energy levels [77]. NV center sensors are based on atomic-scale spin defects in diamond that are highly responsive to magnetic, electric, or thermal changes, and are optically addressable at room temperature [78].

Table 3 compares plasmonic, QD, and NV center-based sensors across key criteria. One important metric is the limit of detection (LOD), which defines the lowest concentration of analyte a sensor can detect. Comparing LODs, NV center-based sensors achieve the highest sensitivity, with LODs reaching the femtomolar range, followed by plasmonic sensors (picomolar) , and then QD-based sensors (nanomolar) [79–81]. Another important metric is multiplexing capability, which refers to the sensor's ability to simultaneously detect multiple targets in a single assay. Among these platforms, QD-based biosensors generally offer superior multiplexing due to their size-tunable, spectrally distinct emission. Plasmonic and NV center sensors can also be multiplexed, but typically with more limited spectral or spatial channels compared to QDs [82].

Table 3. Comparison of Different Quantum Sensing Approaches

| Criterion | Plasmonic | Quantum Dot (QD) | Nitrogen-vacancy (NV) Center |
|---|---|---|---|
| Description | Uses metal nanoparticles to detect molecular interactions via shifts in light absorption or scattering. | Semiconductor nanocrystals that emit fluorescence when excited, with wavelength tunable by their size. | Defects in diamond that detect small magnetic or thermal changes through optical signals. |
| Biosensing Applications | Molecular binding, drug development. | Disease detection, biomarker sensing, bioimaging. | Magnetometry, viral detection, gene sensing, molecular tagging. |
| Sensitivity | Detects refractive index changes with high sensitivity. | Bright, tunable emission supports. | Spin-state readout enables single-molecule detection. |
| Limit of Detection (LOD ↓) | ~ 1 fM–pM | ~ 1 pM–nM | ~ 1 aM–fM |
| Multiplexing | Enabled by engineering distinct plasmonic structures with different resonance frequencies. | Easily achieved through emission tuning across QDs. | Achieved using NVs with varied spin states or spatial encoding. |
| Limitations | Background interference and non-specific binding; requires precise surface engineering. | Fluorescence blinking, toxicity concerns, and spectral instability. | Low photon collection efficiency, need for complex optical setups. |

### 3.2 Quantum Plasmonic Biosensors

Quantum biosensing can overcome classical noise limits through quantum techniques. One powerful implementation of these sensors involves plasmonic biosensors. plasmonic is a field of study focused on the behavior of light at the nanoscale, particularly how it interacts with electrons in metallic nanostructures. Quantum plasmonic biosensors are able to monitor real-time molecular bindings with exceptional sensitivity [83]. They can measure tiny changes at the nanoscale by tracking how electrons tunnel between two closely spaced metal nanoparticles. In 2016, Lerch and Reinhard [12] linked gold nanoparticles by DNA to detect molecular binding events. They showed quantum tunneling could be detected even when the particles are apart in nanometer scale. In this system, quantum tunneling refers to the process by which electrons pass through a nanoscale gap between metal nanoparticles. Table 3 summarizes the main specifications of the described sensor. DNA conductivity measures how well the DNA linker can carry tunneling current, with a reported value of $\sigma_0 = 7.8$ S/m for gaps less than 3 nm. The tunneling current $I$ decreases exponentially with increasing gap distance $S$, as shown in equation 9.

$$I \propto \exp\left(-\frac{2S}{\lambda}\right), \tag{9}$$

where $\lambda$ is the decay length determined by the properties of the barrier (e.g., DNA type). Because of this exponential dependence, even tiny changes in gap size, such as when a protein or ion binds to the DNA lead to measurable changes in the current.

Table 4. Specifications of the discussed quantum plasmonic biosensor

| Parameter | Value |
|---|---|
| Nanoparticle material | Gold (Au) |
| Nanoparticle diameter | 38.5 ± 4.6 nm |
| Interparticle Gap Distance (S) | 0.5–2.8 nm (quantum regime) |
| DNA Linker Length | 40 nucleotide / 80 base pair |
| DNA conductivity | $\sigma_0 = 7.8$ S/m (for S < 3 nm) |

*3.3 Quantum Dot (QD) Biosensors*

QD-based biosensors use semiconductor nanocrystals called quantum dots (QDs) for sensitive detection of biological targets. QDs emit bright, size-tunable light when excited, and their surfaces can be modified to bind specific biomolecules. When a target molecule interacts with a QD, it changes the dot's fluorescence, providing a simple optical readout. Figure 5 shows QD basic structure. QDs are composed of a semiconductor core surrounded by an external shell and coated with surface ligands. The properties of the core primarily determine their optical characteristics, such as light absorption and emission, as well as their semiconducting behavior, including electrical conductivity [84]. Thanks to their strong, stable signal and ability to detect multiple targets at once, QD biosensors are widely used for biomarker detection and molecular imaging in biomedical research.

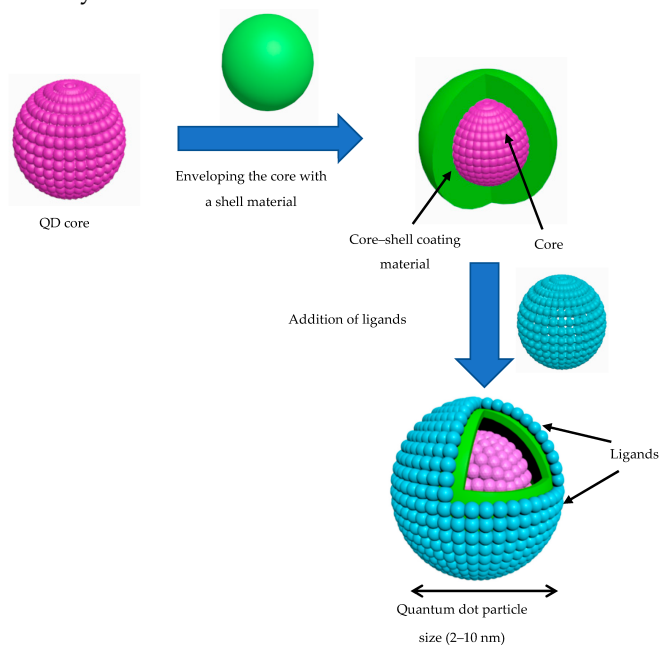

**Figure 5.** Quantum dot basic structure (core, shell, and ligands) [84].

Recent advances demonstrate the versatility of QD-based biosensors across electrochemical, photoelectrochemical, and fluorescence platforms (Table 5). In 2016, Pang et al. [85] introduced a photoelectrochemical biosensor based on NCQD-sensitized perovskite heterojunctions for detecting rheumatoid arthritis-related cells, reaching an LOD as low as 2 cells/mL [81]. In 2018, Roushani et al. [86] developed a graphene quantum dot (GQD)-based electrochemical sensor for streptomycin antibiotic detection, achieving an ultra-low LOD of 0.0033 pg/mL by enhancing electrode conductivity and surface area. In 2020, Muthusankar et al. [87] incorporated nitrogen-doped carbon QDs (NCQDs) with $Co_3O_4$ and carbon nanotubes to sensitively detect anticancer flutamide drugs in urine, to achieve an LOD of 0.0169 µM [78]. Meanwhile, Saadati et al. [88] fabricated a paper-based microfluidic immunosensor using GQD for ovarian cancer biomarker [79]. Moreover, a dual immunosensor by Serafín et al. [89] enabled simultaneous detection of cancer biomarkers using GQD composites for amperometric signal amplification, with LODs of 1.4 and 0.03 ng/mL, respectively [80]. In 2022, Wei et al. [13] reported sulfhydryl-functionalized silicon QDs with 38.5% quantum yield for imaging in live cells and zebrafish, with fluorescence quenching enabling detection down to 13 nM.

*3.4 Nitrogen-Vacancy (NV) Center Diamond Biosensors*

NV center diamond biosensors use atomic-scale defects in diamond for highly sensitive detection of biological signals. The NV center's electron spin state is optically read out and is extremely sensitive to magnetic fields, temperature, or local chemical changes. These biosensors combine quantum-level sensitivity with room-temperature operation and have various applications such as single-molecule detection and nanoscale bioimaging. Figure 6 shows a typical NV center formation in the diamond lattice. It can exist in a neutral ($NV^0$) or negatively charged ($NV^-$) state, with the $NV^-$ state being optically active and sensitive to external fields. A green laser excites the $NV^-$ center in diamond, and the resulting spin-dependent fluorescence is detected by a photodetector. The NV center's spin state is manipulated by microwave radiation, while the local environment, such as magnetic fields or biomolecular interactions shifts the spin resonance [90].

Various studies have used NV centers for highly sensitive biosensing (Table 5). In 2020, Zhang et al. demonstrated a chip-scale NV ensemble magnetometer using pulsed quantum filtering [14]. That same year, Miller et al. introduced a spin-enhanced NV nanodiamond lateral-flow immunosensor for HIV-1 RNA detection, approximately 100,000× more sensitive than conventional gold nanoparticles by separating signal from background through microwave modulation [79]. In 2021, Sharmin et al. applied fluorescent nanodiamonds to track oxidative stress in live cells, using NV spin relaxometry to detect free radicals in real time [91]. Most recently, Li et al. (2022) proposed an NV-based quantum sensor for SARS-CoV-2 RNA detection, employing CRISPR-induced magnetic tags and NV magnetometry for magnetic noise readout, reaching an LOD of a few hundred RNA copies and under 1% false negatives [92].

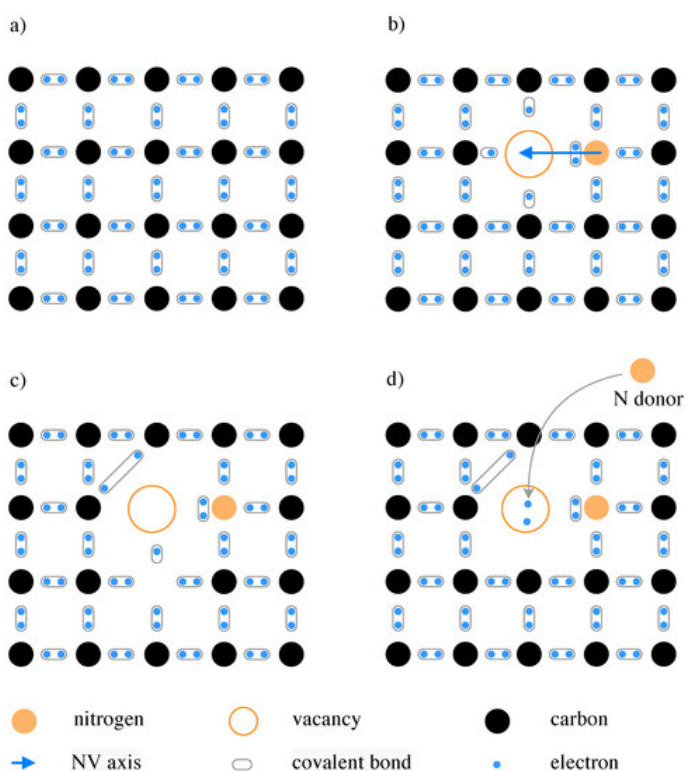

**Figure 6.** Schematic of NV center formation in diamond [90]. **(a)** Substitutional nitrogen atom and nearby vacancy form the NV center. **(b)** Nitrogen bonds to neighbors, leaving a lone pair. **(c)** The NV center in the neutral ($NV^0$) state with five electrons. **(d)** Capture of an extra electron from a nitrogen donor leads to the negatively charged ($NV^-$) state used in quantum sensing.

Table 5. Examples of Quantum Dot (QD) and Nitrogen-Vacancy (NV)-center Diamond Biosensors

| Type | Method | Application | Specification | Working Principle | Ref |
|---|---|---|---|---|---|
| Quantum Dot (QD) Biosensors | GQD electrochemical aptasensor | Streptomycin antibiotics detections | LOD: 0.0033 pg/mL; Linear range: 0.01–812.21 pg/mL | QDs increase electrode conductivity and active area, enabling sensitive detection via electron transfer changes upon target binding. | [86] |
| | NCQD, electrochemical sensor | Anticancer (flutamide) drugs detection in urine | LOD: 0.0169 µM; Linear range: 0.05–590 µM | NCQDs enhance electron transfer and surface area; quantum confinement in CQDs improves electrocatalytic sensing on the modified electrode | [87] |
| | GQD nano-ink immunosensor (Paper-based microfluidic) | Ovarian cancer (Carcinoma Antigen 125) biomarker detection in plasma | LLOQ: 0.001 U/mL; Linear range: 0.001–400 U/mL | Antibody-modified GQD nano-ink enables electrochemical detection via changes in current from antigen–antibody binding | [88] |
| | GQD electrochemical immunosensor (screen-printed electrodes) | Cancer metastasis biomarkers | LOD: 1.4 ng/mL (IL-13R$\alpha$2), 0.03 ng/mL (CDH-17) | GQDs enhance signal via peroxidase-like activity; hybrid nanocarriers amplify current from sandwich immunoassay | [89] |
| | NCQD perovskite Photoelectrochemical | FLS cells detection (Rheumatoid arthritis) | Linear range: $1\times10^4$ – 10 cells/mL; LOD: 2 cells/mL | QDs and perovskite enhance light absorption and charge separation, boosting photocurrent for ultrasensitive detection | [85] |
| | S-SiQD fluorescence probe | ClO$^-$ ion detection, imaging in cells and zebrafish | LOD: 13 nM; Linear range: 0.05–1.8 µM; Quantum yield: 38.5% | Surface sulfhydryl groups enable selective and rapid fluorescence quenching by ClO$^-$ | [13] |
| NV-center Diamond Biosensors | NV-center diamond quantum magnetometer | High-sensitivity magnetic field detection | Sensitivity: 1 nT/√Hz | Optically initializes and reads NV spin states; pulsed protocol filters noise for ultrasensitive magnetometry (biomedical and chip-scale) | [14] |
| | Spin-enhanced NV-center nanodiamond immunosensor | HIV-1 RNA detection (model biotin–avidin) | LOD: $8.2\times10^{-19}$ M; ~100,000× more sensitive than gold-nanoparticles | Microwave-modulated NV center fluorescence separates signal from background, enabling record-low detection limits | [79] |
| | NV center nanodiamond quantum sensor | SARS-CoV-2 RNA detection | LOD: a few hundred RNA copies; <1% false negatives | NV centers in nanodiamonds detect magnetic noise from CRISPR-generated magnetic tags on RNA; optically read out | [92] |
| | NV center fluorescent nanodiamond relaxometry | Detection of free radicals (oxidative stress) in living cells | Single-cell resolution; real-time detection | NV center spin relaxation changes with local magnetic noise from free radicals. | [91] |

**Abbreviations:** QD: Quantum Dots; GQD: Graphene Quantum Dot; NCQD: Nitrogen-doped Carbon Quantum Dot; S-SiQD: Sulfhydryl-functionalized Silicon Quantum Dot; NV: Nitrogen-Vacancy; LOD: Limit of Detection; LLOQ: Lower Limit of Quantification; FLS: Fibroblast-like Synoviocyte; ClO$^-$: Hypochlorite; HIV: Human Immunodeficiency Virus; RNA: Ribonucleic Acid

# 4. On-Chip Quantum Biosensing

*4.1 From Electronic Chips to Photonic Circuits*

Electronic Integrated Circuits (EICs), particularly those based on complementary metal–oxide–semiconductor (CMOS) technology, have long served as the foundation for biosensing devices. In CMOS-based biosensors, signals from biological interactions, such as changes in impedance, voltage, or current are transduced and processed directly on-chip. In addition to CMOS, miniature microelectromechanical systems (MEMS) technologies have advanced the performance of EIC biosensors, offering superior sensitivity, greater miniaturization, and higher power efficiency compared to conventional electronic devices [93–95]. MEMS-based biosensors use mechanical structures, such as cantilevers, membranes, or resonators that respond to forces or mass changes induced by biomolecular interactions [96].

Various electronic integrated circuit-based biosensors have been proposed for diverse biomedical targets (Table 7). For example, Kim et al. [4] developed a bio-impedance IC for cardiovascular monitoring with milliohm sensitivity. Alhoshany et al. [5] reported a CMOS capacitive sensor with femtofarad resolution for cancer biomarkers, and Jang et al. [97] integrated on-chip photodiode arrays for DNA microarray assays. Al Mamun et al. [6] demonstrated amperometric glucose sensing using CMOS potentiostats and carbon nanofibers. Moreover, multiplexed CMOS readout circuits by Li et al. [98] enable the design of low-noise biosensor arrays. In other work, Mehdipoor et al. [99] introduced a MEMS resonator for microfluidic analysis with high mass sensitivity. Mechanical biosensors include Sang et al.'s [100] polydimethylsiloxane (PDMS) membrane sensor for E. coli detection and Kurmendra et al.'s [101] MEMS cantilever arrays for cancer biomarker detection. Timurdogan et al. [102] proposed hepatitis antigen detection using MEMS cantilevers, while Bharati et al. [103] used a ZnO Lamb wave device for DNA biosensing.

Microprocessor chips contain billions of transistors, leading to significant gains in computational speed and efficiency. Despite their advantages, EIC-based biosensors face several fundamental limitations [54]. To bypass CMOS constraints such as interconnect bottlenecks and power dissipation, photonic integrated circuits (PICs) were developed. These optical circuits overcome issues like signal delay and electromagnetic interference, but are still challenged by weak optical nonlinearity, and fabrication complexity [104]. Table 6 compares EIC and PIC biosensors, highlighting their respective strengths and trade-offs. While EICs offer compact integration and cost-effectiveness, PICs provide greater noise immunity and speed. The optimal platform depends on the demands of each application.

**Table 6.** Comparison of electronic and photonic integrated circuit biosensors.

| Technology | Electronic Integrated Circuit (EIC) | Photonic Integrated Circuit (PIC) |
|---|---|---|
| **Advantage** | Compact, scalable, cost-effective, readily available | High speed, low power consumption, immunity to electromagnetic interference, support high levels of multiplexing. |
| **Limitation** | Susceptible to electrical noise, electrode drift, and electromagnetic interference, speed limitation. | Complex fabrication, need precise optical alignment, technology is still evolving. |
| **Specifications** | SNR: 15–30 dB<br>Multiplexing: up to 100 channels<br>Footprint: ~0.01–10 mm² | SNR: 30–50 dB<br>Multiplexing: 50–1000 channels<br>Footprint: ~1–100 mm² |
| **Application** | Glucose biosensors, protein assays, cardiac sensors | Nucleic acid detection, protein biomarker panels, virus assays |

**Abbreviations:** SNR: Signal-to-Noise Ratio.

14 of 26**Table 7.** Comparison of some electronic integrated circuit (EIC)-based biosensors

| Type | Method | Working Principle | Specification | Application | Ref |
|------|--------|-------------------|---------------|-------------|-----|
| CMOS | Bio-impedance IC | Impedance-based voltage/current sensing | Supply: 0.5 V; Power: <10 μW; Noise: 15.28 mΩ/√Hz; Phase error: <1° | Cardiovascular disease | [4] |
| | Capacitive CMOS | Capacitance change due to biomolecular binding | Power: 2.1 μW; Capacitance range: 16.137 pF; Resolution: 4.5 fF | Cancer enzyme biomarker (oncology) | [5] |
| | CMOS fluorescence microarray | Fluorescent emission, photodiode readout | Excitation: 532 nm; Dark current: ~12 fA; ADC resolution: 14-bit; Array size: 16×16 | Genomics, DNA hybridization (bio-assay platforms) | [97] |
| | CMOS potentiostat with carbon nanofiber | Amperometric current changes from glucose oxidation | Power: 71.7 μW; Sensitivity: 50–200 nA/mM; Electrode area: 0.09 mm²; Detection range: 0.5–7 μA | Diabetes, glucose | [6] |
| | CMOS picoamp current readout | Low-noise current readout for electrochemical biosensor arrays | Noise: 7.2 pA_rms; Power: 21 μW/channel; Area: 0.06 mm²/channel; Bandwidth: 11.5 kHz | Multiplexed biosensor readout, DNA sequencing | [98] |
| MEMS | MEMS resonator + microfluidics | Resonant frequency shift due to particle mass | Frequency: 16.5 kHz; Displacement: 1.44 μm; Q-factor: 49; Sensitivity: 1×10$^{11}$ Hz/kg | Digital microfluidics, droplet (lab-on-chip) | [99] |
| | PDMS membrane | Membrane deflection from surface stress, interferometric readout | Membrane: 2.5×2.5 mm; Thickness: 35 μm; Young's modulus: 12 kPa–2.5 MPa; Sensitivity: 0.56×10$^{-5}$ N/m | E. coli (microbiology, pathogen screening) | [100] |
| | MEMS cantilever | Adsorption-induced stress, piezoresistive readout | LOD: 1 ppb (VOC); Cantilever: 150×500 μm; Resonant freq: ~12 kHz; Response time: <10 s | Cancer (lung, breast, prostate) | [101] |
| | MEMS cantilever | Resonant frequency shift of nickel cantilever | LOD: 0.1 ng/mL; Dynamic range: >1000×; Array: up to 16 cantilevers; Detection time: ~20 s | Hepatitis A/C virus, serum analysis | [102] |
| | MEMS ZnO Lamb wave resonator | Piezoelectric Lamb wave, frequency shift by DNA mass | Sensitivity: 310 Hz/ng/μL (DNA); LOD: 82 pg/μL; Frequency: 137 MHz; Membrane: 4.5×5.9 mm | DNA biosensing (meningitis pathogen) | [103] |

**Abbreviations:** IC: integrated circuit; ADC: analog-to-digital converter; PDMS: polydimethylsiloxane; Q-factor: quality factor; LOD: limit of detection; CMOS: complementary metal–oxide–semiconductor; MEMS: miniature microelectromechanical systems; ZnO: zinc oxide; DNA: deoxyribonucleic acid; E. coli: Escherichia coli.

*4.2 Photonic Integrated Circuits for Biosensing*

Photonic Integrated Circuits (PICs) process information using light instead of electrical signals. Recently, silicon photonic components are considered among the most promising platforms for photonic integration. They combine high refractive index contrast with full compatibility to CMOS microelectronic fabrication, and PICs can take advantage of established microelectronics manufacturing techniques (Figure 7c) [105]. PIC-based biosensors are commonly waveguide-based sensors, using ring resonators or interferometers to monitor changes in refractive index with high sensitivity. Changes in refractive index are typically quantified using the refractive index unit (RIU), which is as a standard metric for detection performance. RUI is a standard unit which measures how much light slows down as it passes through a material compared to vacuum. This unit allows for precise reporting of how sensitive a device is to small variations in the optical properties of the sample [106].

A waveguide is a narrow optical channel that directs light along a defined path on the chip. Some of the light traveling in the waveguide extends just outside its surface, forming what is called an evanescent field. When target biomolecules attach to the surface of the waveguide, they interact with this evanescent field and cause a change in the local refractive index. This, in turn, alters the properties of the light traveling through the waveguide, such as its speed or phase. Among the most widely used waveguide-based biosensing structures are ring resonators and Mach–Zehnder interferometers (MZIs), both of which exploit the evanescent field to detect local refractive index changes near the waveguide surface [107]. A ring resonator consists of a tiny closed-loop optical waveguide that traps light. It allows only specific wavelengths to resonate constructively. When biomolecules attach to the ring's surface, they change the local refractive index, causing a shift in the resonance wavelength. An MZI splits light into two separate paths. One path is exposed to the sample, so that biomolecular binding alters its refractive index and therefore its optical path length. When the two light beams recombine, any difference in phase between the paths results in constructive or destructive interference, which can be measured as a change in output intensity. MZIs have demonstrated sensitivity to refractive index changes as small as $10^{-7}$ RIU [108].

Various PIC biosensors based on MZIs, and ring resonators have been proposed for label-free biosensing (Table 8). For example, Vogelbacher et al. [109] developed a $Si_3N_4$ MZI biosensor with an integrated laser diode for streptavidin–biotin assays (Figure 7d). Densmore et al. [110] demonstrated a spiral photonic wire MZI for antibody–antigen reactions. Crespi et al. [111] introduced a 3D femtosecond-laser-written MZI enabling spatially resolved label-free sensing with a 10 μm resolution. A rib-waveguide MZI was reported by Prieto et al. [8], for water pollutant monitoring. In micro-ring platforms, Ning et al. [9] proposed a resonator for femtomolar SARS-CoV-2 detection. Bryan et al. [10] realized multiplexed antibody profiling using SiN ring arrays (Figure 7a, Figure 7b). Kumar et al. [112] developed a slotted plasmonic ring resonator for water pollutant sensing. Singh et al. [113] presented a rectangular semi-ring waveguide biosensor for hepatitis detection. Haron et al. [114] integrated a tunable photonic crystal ring with a PN phase shifter for dynamic wavelength control. Finally, Voronkov et al. [115] proposed a microring with dual-ring interrogation for refractometric liquid sensing (Figure 7e).

**Table 8.** Comparison of some photonic integrated circuit (PIC)-based biosensors

| Type | Method | Working Principle | Specification | Application | Ref |
|---|---|---|---|---|---|
| Mach–Zehnder Interferometer (MZI) | $Si_3N_4$ MZI + integrated laser source | Optical phase shift between split paths caused by binding events | Sensitivity: $6.8 \times 10^{-6}$ RIU; LOD: ~1 ng/mL; Footprint: $3.5 \times 0.6$ mm² | Streptavidin–biotin sensing | [109] |
| | Spiral photonic wire MZI in SOI | Compact spiral MZI maximizes evanescent field–analyte interaction | LOD: 0.25 pg/mm²; Waveguide: 1.8 mm spiral; Channel width: 190 μm | Antibody–antigen detection | [110] |
| | 3D femtosecond-laser-written MZI | Vertical sensing arm intersects fluidic channel orthogonally | LOD: $1 \times 10^{-4}$ RIU; Spatial resolution: 10 μm; Channel width: 150 μm | Resolved label-free sensing | [111] |
| | $Si_3N_4$ rib nanodevice MZI | Evanescent field senses refractive index shifts | LOD: $7 \times 10^{-6}$ RIU; Sensor length: 15 mm; Core thickness: 250 nm | Water pollutants detection | [8] |
| Ring Resonator | Subwavelength grating micro-ring | Antigen–antibody binding alters resonance in the ring | LOD: 1.31 fM; Detection time: 15 min; Q-factor: ~30,000 | SARS-CoV-2 & influenza | [9] |
| | SiN ring resonator array with photonic packaging | Parallel resonance shift in functionalized rings | Q-factor: $>4 \times 10^4$; ER: >20 dB; FSR: 2.54 nm | Respiratory antibody profiling | [10] |
| | Slotted plasmonic ring resonator | Enhanced field in slot increases sensitivity to refractive index | Sensitivity: 1609 nm/RIU; Spectral shift: 29.6 nm (ΔRI = 0.0184); Slot width: 10 nm | Water pollutants detection | [112] |
| | Rectangular semi-ring optical waveguide biosensor | Refractive index changes from biomolecular interaction shifts | LOD: 0.12 ng/mL; Linear range: 0.1–100 ng/mL; Sensor width: 800 nm | Hepatitis B virus | [113] |
| | Photonic crystal ring + PN phase shifter | Phase-tunable ring enables reconfigurable resonance sensing | Tuning range: 1.3–1.7 μm; Ring radius: 6 μm; Phase shift: π | General biomedical sensing | [114] |
| | Racetrack micro-ring resonator with dual ring readout | Refractive index changes shift resonance; monitored by intensity changes via secondary ring | Sensitivity: 110 nm/RIU; Delay: 10 ns at 100 MHz; Fabrication tolerance: ±8 nm | Liquid refractometry | [115] |

**Abbreviations:** MZI: Mach–Zehnder Interferometer; PIC: Photonic Integrated Circuit; RIU: Refractive Index Unit; LOD: Limit of Detection; SOI: Silicon-On-Insulator; $Si_3N_4$: Silicon Nitride; Q-factor: Quality Factor; ER: Extinction Ratio; FSR: Free Spectral Range PN: P-type/N-type Semiconductor Junction.

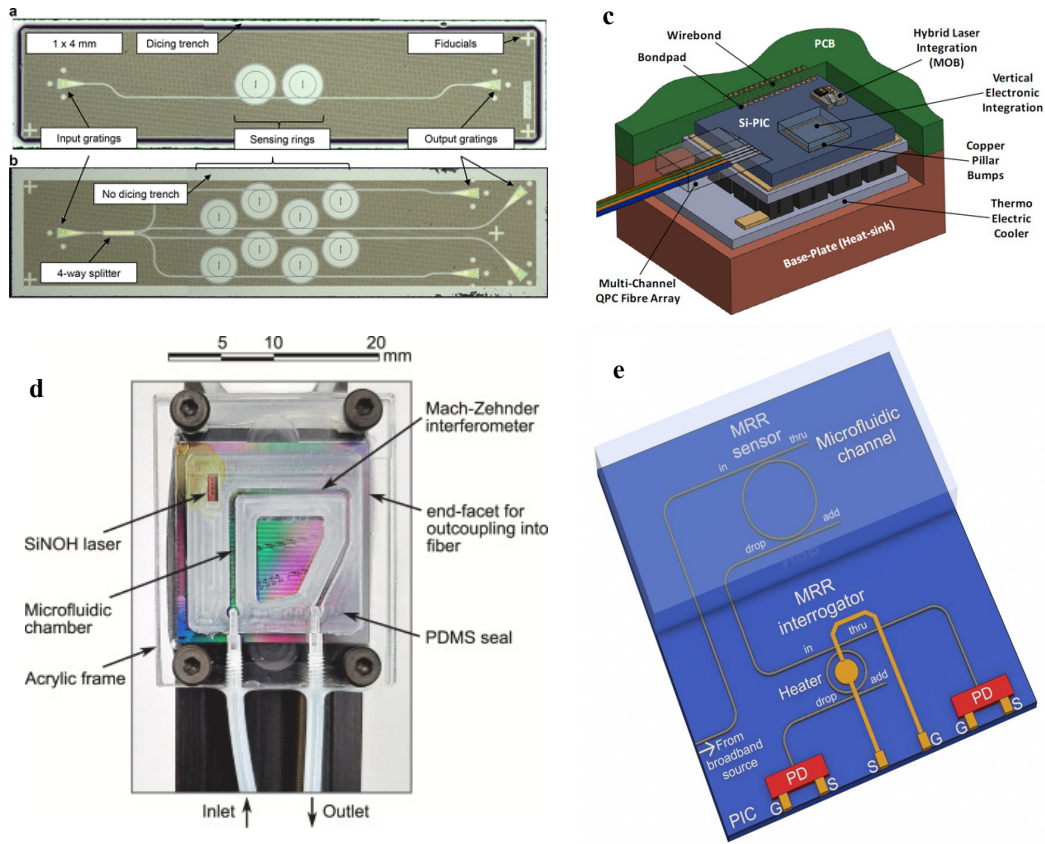

**Figure 7.** Examples of photonic integrated circuit (PIC)-based biosensors **(a)** Fabricated and diced singleplex PIC for antibody profiling in upper respiratory disease, by Bryan et al. [10] **(b)** Fabricated and diced multiplex PIC, by Bryan et al. [10] **(c)** Schematic of a Silicon PIC packaged with a multi-channel quasi-planar coupled (QPC) fiber-array, a hybrid-integrated laser source based on a micro-optic bench (MOB), a vertically integrated electronic integrated circuit (EIC), and a thermo-electric cooler. [116] **(d)** Implementation of the waveguide MZI photonic lab-on-a-chip biosensor, by Vogelbacher et al. [109] **(e)** 3D design draft of the integrated microring (MRR)-based Photonic Sensing System for Liquid Refractometry, by Voronkow et al. [115].

### 4.3 Building Toward Integrated Quantum Photonics (IQPs)

Chip-scale quantum system architectures integrate quantum elements directly onto a single microelectronic chip using advanced semiconductor fabrication. To build compact integrated quantum photonic (IQP) chips, silicon-based approaches have been widely investigated. Despite progress on individual components, achieving full integration likely demands the use of multiple materials. So far, a chip-based IQP remains a scientific challenge. However, with successful integration of on-chip light sources, a fully integrated silicon based IQP holds strong potential for future microelectronic [117]. Lithium niobate ($LiNbO_3$) is a material used in integrated photonics, along with diamond, silicon carbide (SiC), and III–V semiconductors such as Gallium arsenide (GaAs), Indium phosphide (InP), Gallium nitride (GaN). Silicon nanostructures, in particular, are now recognized as a leading platform for quantum photonic technology [118].

IQP devices, including quantum light sources, phase shifters, and single-photon detectors, form the essential building blocks for on-chip quantum information processing and sensing. Among integrated quantum photonic devices, quantum light sources and phase shifters are already compatible with microelectronic fabrication and can be integrated on silicon chips. Single-photon detectors (SPDs) can also be integrated, especially single-photon avalanche diodes (SPADs), though superconducting nanowire single-photon detectors (SNSPDs) often require cryogenic cooling, which limits their practical use.

Each of IQP components contributes distinct functionalities to circuits, and their roles can be better understood by examining them individually. Quantum light sources generate non-classical states of light, such as single photons or entangled photon pairs. These devices often rely on nonlinear optical processes, where one photon is converted into two lower-energy photons. The main advantage of quantum light sources is their foundational role in enabling secure communication and quantum computation [119]. For instance, Politi et al. [120] designed a silicon chip integrating single-photon sources, beam splitters and phase shifters, and single-photon detectors. The device used micrometer-scale waveguides, thermo-optic phase shifters, and achieved low optical losses (<1 dB/cm).

Once photons are generated, their paths and phases must be precisely manipulated. This task is performed by phase shifters and universal interferometric circuits. Phase shifters and universal circuits are the core components for implementing quantum gates. These devices manipulate the paths and phases of single photons and applies a unitary transformation described by equation 10, allowing precise control over photon interference. Their universality makes them critical for scalable optical quantum computing [121]. For example, Grottke et al. [122] designed a MEMS-based phase shifter on silicon nitride ($Si_3N_4$) that uses electrostatic actuation to control phase with low loss (<0.5 dB), low voltage ($V\pi$ = 2 V), and short length (210 μm). The device achieved up to $13.3\pi$ phase shift and MHz operation. The device offers advantages such as low power consumption, compact size, and high modulation speed.

$$U(\theta) = \begin{pmatrix} 1 & 0 \\ 0 & e^{i\theta} \end{pmatrix}. \qquad (10)$$

Finally, the detection stage completes the quantum photonic system by converting single-photon events into measurable electrical outputs. Single-photon detectors (SPDs) are designed to register the arrival of individual photons, converting photonic events into electrical signals. SPDs are crucial for quantum measurement, but they may require cryogenic operation. Common implementations include superconducting nanowire single-photon detectors (SNSPDs) and single-photon avalanche diodes (SPADs) [123]. SPAD and SNSPD technologies rely on fundamentally different detection mechanisms. Avalanche photodetectors are highly sensitive semiconductor photodiodes that utilize the photoelectric effect to convert incident photons into electrical signals. When a photon is absorbed, it creates an electron–hole pair within the semiconductor, initiating charge multiplication. SPADs operate in Geiger mode, where the device is biased above its breakdown voltage. In this regime, absorption of a single photon can trigger an avalanche current, which persists until the bias is reduced below breakdown. The avalanche process enables detection of individual photons [124]. As an example, Ref. [125] presents a room-temperature, CMOS-compatible Ge–Si SPAD array with 32 × 32 pixels, each with a diameter of 15 μm. The device achieves a photon detection probability of up to 12% at 1,310 nm. The breakdown voltage is approximately 10.3 V. The SPAD array enabled high-resolution three-dimensional time-of-flight imaging, demonstrating the platform's promise for quantum photonic and imaging applications. In contrast, SNSPDs exploit superconductivity to achieve even higher detection efficiency, albeit under cryogenic conditions. Superconducting nanowire single-photon detectors (SNSPDs) operate by cooling a nanowire below its critical temperature so that electrons form a superconducting state. When a photon is absorbed, it creates a localized hotspot that disrupts superconductivity, forcing the nanowire to briefly switch to a normal state and generating a detectable voltage pulse [126]. In 2024, Hao et al. [127] developed a compact 2×2-pixel SNSPD array using 5 nm niobium nitride (NbN) nanowires for high-speed optical communications. Operating at 1.5 K, the detector achieved 91.6% efficiency. The device supports photon-number resolution up to 24 photons. In system tests, it maintained error-free data transmission at 0.8 noise photons/slot and up to 1.5 Gbps. While offering high sensitivity and dynamic range, the array requires cryogenic cooling.

Despite these advances, many challenges remain in implementing fully integrated quantum photonic sensors. For instance, an innovative design [128] using asymmetric waveguides for bidirectional transmission on printed circuit boards (PCBs) offers a potential pathway to connect IQP with conventional microelectronics, indicating that much work is still needed in this growing field [129]. In July 2025, Kramnik et al. [15] demonstrated the first commercial integrated quantum photonic chip that combines electronic, photonic, and quantum components on a single silicon platform, which is a major step toward IQP biosensors (Figure 9).

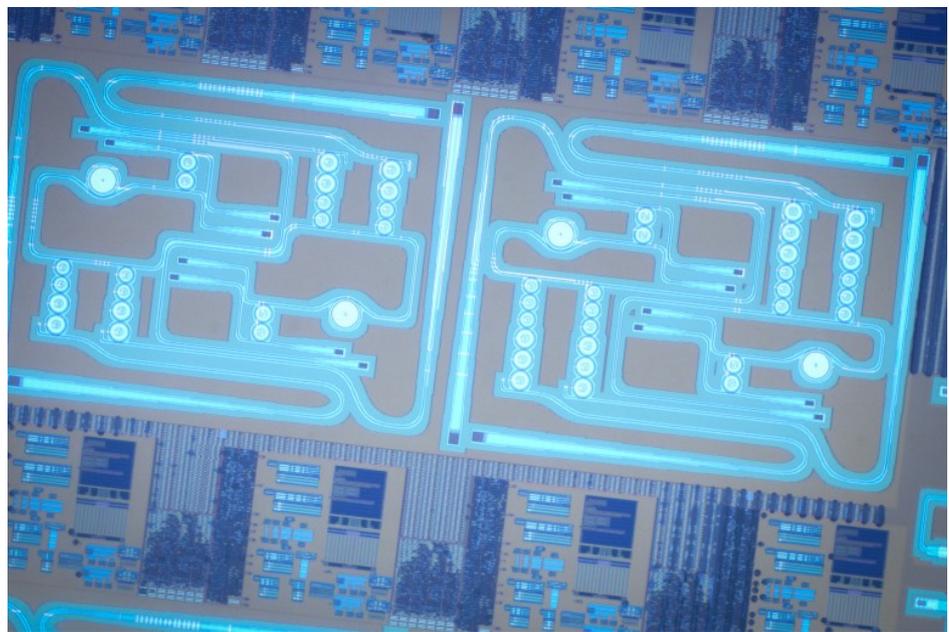

**Figure 8.** First time electronic-photonic quantum chip, manufactured by Kramnik et al. [15] The chip can stabilize itself despite temperature changes and fabrication variations, which is an essential requirement for scaling up quantum systems. It also doesn't need for large external equipment. (Image source: https://news.northwestern.edu/stories/2025/07/first-electronic-photonic-quantum-chip-manufactured-in-commercial-foundry/ [date accessed: 30 July 2025]).

## 5. Research Challenges and Future Perspectives

Quantum biosensors, electronic integrated circuits (EICs), and photonic integrated circuits (PICs) each offer distinct strengths yet face limitations. Quantum biosensors, despite their exceptional sensitivity, often face issues such as quantum state decoherence and the complexity of integration onto compact chip-scale devices. Similarly, EIC-based biosensors provide cost-effective and compact solutions; however, they encounter fundamental constraints, including electrical noise susceptibility, interconnect bottlenecks, and challenges in scaling signal multiplexing. PIC-based biosensors effectively address these electrical limitations through their immunity to electromagnetic interference and ability to achieve high multiplexing and low-power operation. Yet, they continue to struggle with complicated fabrication processes, optical alignment, and nonlinearities.

Advancing towards integrated quantum photonics (IQP) for biosensing is a promising way to address these limitations. Recent steps toward IQP integration have demonstrated the feasibility of on-chip quantum sources, universal quantum gates, and single-photon detectors. Nevertheless, current IQP circuits face notable challenges, including maintaining quantum coherence at room temperature, efficient photon coupling, minimizing optical losses, and achieving reliable quantum state readout without extensive cryogenic infrastructure. Innovations such as silicon-compatible quantum emitters, MEMS-enabled dynamic phase modulators, and CMOS-integrated single-photon avalanche diodes (SPADs) are mitigating these constraints. For instance, room-temperature materials like diamond NV centers and Si-based SPAD arrays can significantly reduce system complexity and enhance operational robustness. Ultimately, the trajectory toward practical IQP biosensors will require ground-breaking solutions. Future progress may arise from adopting hybrid-material platforms, such as combining silicon carbide (SiC) or lithium niobate ($LiNbO_3$) waveguides, known for their superior nonlinear optical properties and room-temperature quantum coherence, with silicon photonics to boost photon coupling efficiency and quantum state stability. Integrating two-dimensional quantum materials like hexagonal boron nitride (hBN) or transition metal dichalcogenides (TMDs) might further provide tunable quantum emitters that facilitate on-chip generation and manipulation of entangled photons at room temperature.

To resolve challenges with photon detection and signal loss, next-generation quantum sensors could utilize superconducting nanowire detectors embedded within hybrid photonic-electronic layers, thermally isolated by advanced micro-scale cryocoolers. Alternatively, meta-surfaces or photonic crystal structures designed specifically to enhance photon extraction efficiency may dramatically improve the quantum sensing even at ambient temperatures. Quantum-classical interfacing issues could be creatively managed by embedding quantum states directly into topological photonic circuits. Moreover, future manufacturing of IQP biosensors might move toward three-dimensional integration, incorporating quantum, photonic, and electronic components vertically stacked through advanced wafer-bonding and through-silicon-via (TSV) techniques. Such approaches would accelerate mass production and standardization, crucial for real-world clinical adoption. Envisioning this integrated quantum-photonic roadmap illustrates a promising future where IQP biosensors provide diagnostic sensitivity in portable, and universally accessible healthcare technologies.

## 6. Conclusions

Quantum biosensors offer notable advantages over classical sensing platforms, particularly in achieving fast speed, high sensitivity, single-molecule resolution, and low-noise performance. This review compared different quantum sensors and evaluated their integration with microelectronic and photonic technologies. Electronic integrated circuits (EICs) and photonic integrated circuits (PICs) have advanced on-chip biosensing, yet both encounter limitations related to scaling, fabrication, and signal interference. Integrated quantum photonics (IQP) introduces a viable path to address these issues through discrete quantum states and enhanced measurement precision. Despite early-stage progress, current IQP devices face practical challenges in material uniformity, optical loss, and device complexity. Addressing these gaps will require new device layouts, better packaging strategies, and more compatible fabrication flows with silicon photonics. Continued progress in these directions can turn quantum biosensing from a laboratory concept into a clinically and industrially practical solution in future.

# References

1. Sasidhar, B. ADVANCEMENTS IN BIOSENSOR TECHNOLOGIES: FROM NANOBIOSENSORS TO BIOCOMPATIBLE AND OPTICAL SYSTEMS FOR CLINICAL AND ENVIRONMENTAL APPLICATIONS. *World Journal of Pharmaceutical Sciences* **2025**.
2. Hassan, M.M.; Xu, Y.; Zareef, M.; Li, H.; Chen, Q. Recent Progress in Chemometrics Driven Biosensors for Food Application. *TrAC Trends in Analytical Chemistry* **2022**, *156*, 116707, doi:10.1016/j.trac.2022.116707.
3. Athanassov, A.S. THE LITTLE TRANSISTOR AND THE CCAS REVOLUTION. **2007**.
4. Kim, K.; Kim, J.-H.; Gweon, S.; Kim, M.; Yoo, H.-J. A 0.5-V Sub-10-μW 15.28-mΩ/√Hz Bio-Impedance Sensor IC With Sub-1° Phase Error. *IEEE Journal of Solid-State Circuits* **2020**, *55*, 2161–2173, doi:10.1109/JSSC.2020.2991511.
5. Alhoshany, A.; Sivashankar, S.; Mashraei, Y.; Omran, H.; Salama, K.N. A Biosensor-CMOS Platform and Integrated Readout Circuit in 0.18-Mm CMOS Technology for Cancer Biomarker Detection. *Sensors* **2017**, *17*, 1942, doi:10.3390/s17091942.
6. Al Mamun, K.A.; Islam, S.K.; Hensley, D.K.; McFarlane, N. A Glucose Biosensor Using CMOS Potentiostat and Vertically Aligned Carbon Nanofibers. *IEEE Transactions on Biomedical Circuits and Systems* **2016**, *10*, 807–816, doi:10.1109/TBCAS.2016.2557787.
7. Nagarajan, R.; Doerr, C.; Kish, F. *Optical Fiber Telecommunications VIA: Chapter 2. Semiconductor Photonic Integrated Circuit Transmitters and Receivers*; Elsevier Inc. Chapters, 2013; ISBN 978-0-12-806056-8.
8. Prieto, F.; Sepúlveda, B.; Calle, A.; Llobera, A.; Domínguez, C.; Abad, A.; Montoya, A.; Lechuga, L.M. An Integrated Optical Interferometric Nanodevice Based on Silicon Technology for Biosensor Applications. *Nanotechnology* **2003**, *14*, 907, doi:10.1088/0957-4484/14/8/312.
9. Ning, S.; Chang, H.-C.; Fan, K.-C.; Hsiao, P.-Y.; Feng, C.; Shoemaker, D.; Chen, R.T. A Point-of-Care Biosensor for Rapid Detection and Differentiation of COVID-19 Virus (SARS-CoV-2) and Influenza Virus Using Subwavelength Grating Micro-Ring Resonator. *Applied Physics Reviews* **2023**, *10*, doi:10.1063/5.0146079.
10. Bryan, M.R.; Butt, J.N.; Ding, Z.; Tokranova, N.; Cady, N.; Piorek, B.; Meinhart, C.; Tice, J.; Miller, B.L. A Multiplex "Disposable Photonics" Biosensor Platform and Its Application to Antibody Profiling in Upper Respiratory Disease. *ACS Sens.* **2024**, *9*, 1799–1808, doi:10.1021/acssensors.3c02225.
11. NASA-Industry Team Creates and Demonstrates First Quantum Sensor for Satellite Gravimetry - NASA 2018.
12. Lerch, S.; Reinhard, B.M. Quantum Plasmonics: Optical Monitoring of DNA-Mediated Charge Transfer in Plasmon Rulers. *Advanced Materials* **2016**, *28*, 2030–2036, doi:10.1002/adma.201503885.
13. Wei, N.; Sun, Y.-C.; Guo, X.-F.; Wang, H. Synthesis of Sulfhydryl Functionalized Silicon Quantum Dots with High Quantum Yield for Imaging of Hypochlorite in Cells and Zebrafish. *Microchim Acta* **2022**, *189*, 329, doi:10.1007/s00604-022-05435-x.
14. Zhang, Y.; Li, Z.; Feng, Y.; Guo, H.; Wen, H.; Tang, J.; Liu, J. High-Sensitivity DC Magnetic Field Detection with Ensemble NV Centers by Pulsed Quantum Filtering Technology. *Opt. Express, OE* **2020**, *28*, 16191–16201, doi:10.1364/OE.392279.
15. Kramnik, D.; Wang, I.; Ramesh, A.; Fargas Cabanillas, J.M.; Gluhović, Đ.; Buchbinder, S.; Zarkos, P.; Adamopoulos, C.; Kumar, P.; Stojanović, V.M.; et al. Scalable Feedback Stabilization of Quantum Light Sources on a CMOS Chip. *Nat Electron* **2025**, *8*, 620–630, doi:10.1038/s41928-025-01410-5.
16. Hughes, C.; Isaacson, J.; Perry, A.; Sun, R.F.; Turner, J. What Is a Qubit? In *Quantum Computing for the Quantum Curious*; Hughes, C., Isaacson, J., Perry, A., Sun, R.F., Turner, J., Eds.; Springer International Publishing: Cham, 2021; pp. 7–16 ISBN 978-3-030-61601-4.
17. Robert, A.; Barkoutsos, P.K.; Woerner, S.; Tavernelli, I. Resource-Efficient Quantum Algorithm for Protein Folding. *npj Quantum Inf* **2021**, *7*, 38, doi:10.1038/s41534-021-00368-4.
18. Sajjan, M.; Li, J.; Selvarajan, R.; Sureshbabu, S.H.; Kale, S.S.; Gupta, R.; Singh, V.; Kais, S. Quantum Machine Learning for Chemistry and Physics. *Chem. Soc. Rev.* **2022**, *51*, 6475–6573, doi:10.1039/D2CS00203E.
19. Sawaya, N.P.D.; Huh, J. Quantum Algorithm for Calculating Molecular Vibronic Spectra. *J. Phys. Chem. Lett.* **2019**, *10*, 3586–3591, doi:10.1021/acs.jpclett.9b01117.
20. Huh, J.; Guerreschi, G.G.; Peropadre, B.; McClean, J.R.; Aspuru-Guzik, A. Boson Sampling for Molecular Vibronic Spectra. *Nature Photon* **2015**, *9*, 615–620, doi:10.1038/nphoton.2015.153.
21. Barr, A.J.; Fabbrichesi, M.; Floreanini, R.; Gabrielli, E.; Marzola, L. Quantum Entanglement and Bell Inequality Violation at Colliders. *Progress in Particle and Nuclear Physics* **2024**, *139*, 104134, doi:10.1016/j.ppnp.2024.104134.

22. Chaudhuri, B.; Chaudhuri, B. Application of Quantum Entanglement to Explain the Healing Mechanism by Highly Diluted Homoeopathic Medicines. *Indian Journal of Research in Homoeopathy* **2025**, *19*, 3–14, doi:10.53945/2320-7094.1963.
23. He, M. Entanglement-Enhanced Bioimaging and Sensing, California Institute of Technology, 2024.
24. André Michaud Critical Analysis of the Origins of Heisenberg's Uncertainty Principle. *Journal of Modern Physics* **2024**, *15*, 765–795, doi:10.4236/jmp.2024.156034.
25. Gurjar, V.; Rajan, A.; Chaturvedi, A.; Tiwari, R.; Ratre, P.; Mishra, P. Deep Learning-Enabled Quantum Imaging: Future-Ready Nanosensing Technologies for Preventive Health Interventions. *Computational and Structural Biotechnology Reports* **2025**, *2*, 100053, doi:10.1016/j.csbr.2025.100053.
26. Ghamsari, M.S.; Baniasadi, F. Quantum for Biology: Spectroscopy and Sensing. *Innov. Emerg. Technol.* **2024**, *11*, doi:10.1142/s2737599424300058.
27. Mauranyapin, N.P.; Terrasson, A.; Bowen, W.P. Quantum Biotechnology. *Advanced Quantum Technologies* **2022**, *5*, 2100139, doi:10.1002/qute.202100139.
28. Zheng, W.; Wang, H.; Schmieg, R.; Oesterle, A.; Polzik, E.S. Entanglement-Enhanced Magnetic Induction Tomography. *Phys. Rev. Lett.* **2023**, *130*, 203602, doi:10.1103/PhysRevLett.130.203602.
29. Castillo, J.C.R. Differential Equations: Fundamentals, Solution Methods, and Applications in Dynamical Systems and Chaos Theory. *Ibero Ciencias - Revista Científica y Académica - ISSN 3072-7197* **2025**, *4*, 22–42, doi:10.63371/ic.v4.n2.a36.
30. Dal Lin, C.; Romano, P.; Iliceto, S.; Tona, F.; Vitiello, G. On Collective Molecular Dynamics in Biological Systems: A Review of Our Experimental Observations and Theoretical Modeling. *International Journal of Molecular Sciences* **2022**, *23*, 5145, doi:10.3390/ijms23095145.
31. Matarèse, B.F.E.; Purushotham, A. Quantum Oncology. *Quantum Reports* **2025**, *7*, 9, doi:10.3390/quantum7010009.
32. Sutcliffe, M.J.; Scrutton, N.S. Enzymology Takes a Quantum Leap Forward. *Philos Trans A Math Phys Eng Sci* **2000**, *358*, 367–386, doi:10.1098/rsta.2000.0536.
33. Niazi, S.K. The Quantum Paradox in Pharmaceutical Science: Understanding Without Comprehending—A Centennial Reflection. *Int J Mol Sci* **2025**, *26*, 4658, doi:10.3390/ijms26104658.
34. Slocombe, L.; Sacchi, M.; Al-Khalili, J. An Open Quantum Systems Approach to Proton Tunnelling in DNA. *Commun Phys* **2022**, *5*, 109, doi:10.1038/s42005-022-00881-8.
35. Löwdin, P.-O. Proton Tunneling in DNA and Its Biological Implications. *Rev. Mod. Phys.* **1963**, *35*, 724–732, doi:10.1103/RevModPhys.35.724.
36. Ali, A.; Naeem, M.Y.; Selamoglu, Z.; Naqvi, M.R. Exploring Quantum in Cancer Biology: A Comprehensive Review of Nontrivial Quantum Events. *Archives of Razi Institute* **2025**, *80*.
37. Nicolini, C.; Sivozhelezov, V. Quantum Effects and Genetic Code: Dynamics and Information Transfer in DNA Replication S. Mayburov.
38. Adams, B.; Sinayskiy, I.; van Grondelle, R.; Petruccione, F. Quantum Tunnelling in the Context of SARS-CoV-2 Infection. *Sci Rep* **2022**, *12*, 16929, doi:10.1038/s41598-022-21321-1.
39. Haseeb, M.W.; Toutounji, M. Vibration Assisted Electron Tunnelling in COVID-19 Infection Using Quantum State Diffusion. *Sci Rep* **2024**, *14*, 12152, doi:10.1038/s41598-024-62670-3.
40. Aslam, N.; Zhou, H.; Urbach, E.K.; Turner, M.J.; Walsworth, R.L.; Lukin, M.D.; Park, H. Quantum Sensors for Biomedical Applications. *Nat Rev Phys* **2023**, *5*, 157–169, doi:10.1038/s42254-023-00558-3.
41. Schofield, H.; Boto, E.; Shah, V.; Hill, R.M.; Osborne, J.; Rea, M.; Doyle, C.; Holmes, N.; Bowtell, R.; Woolger, D.; et al. Quantum Enabled Functional Neuroimaging: The Why and How of Magnetoencephalography Using Optically Pumped Magnetometers. *Contemporary Physics* **2022**, *63*, 161–179, doi:10.1080/00107514.2023.2182950.
42. Goryanin, I.I.; Damms, B.; Vesnin, S.; Shevelev, O.; Gorya, I. Exploring the Interface of Microwave Technology, Quantum Computing and Neuroscience 2024.
43. D. Allert, R.; D. Briegel, K.; B. Bucher, D. Advances in Nano- and Microscale NMR Spectroscopy Using Diamond Quantum Sensors. *Chemical Communications* **2022**, *58*, 8165–8181, doi:10.1039/D2CC01546C.
44. Georgiev, D.D. Quantum Information in Neural Systems. *Symmetry* **2021**, *13*, 773, doi:10.3390/sym13050773.
45. Parmar, S.J.; Parmar, V.R.; Verma, J.; Roy, S.; Bhattacharya, P. Quantum Computing: Exploring Superposition and Entanglement for Cutting-Edge Applications. In Proceedings of the 2023 16th International Conference on Security of Information and Networks (SIN); November 2023; pp. 1–6.

46. Salloum, H.; Lukin, R.; Mazzara, M. Quantum Computing in Drug Discovery: A Review of Quantum Annealing and Gate-Based Approaches.; September 20 2024.
47. Chow, J.C.L. Quantum Computing in Medicine. *Medical Sciences* **2024**, *12*, 67, doi:10.3390/medsci12040067.
48. Singh, N.; Pokhrel, S.R. Modeling Quantum Machine Learning for Genomic Data Analysis 2025.
49. Wang, Y.; Zhou, X. Efficient Quantum Algorithm for Lattice Protein Folding. *Quantum Sci. Technol.* **2025**, *10*, 015056, doi:10.1088/2058-9565/ada08e.
50. Kiani, B.T.; Villanyi, A.; Lloyd, S. Quantum Medical Imaging Algorithms 2020.
51. Takeda, K.; Noiri, A.; Nakajima, T.; Kobayashi, T.; Tarucha, S. Quantum Error Correction with Silicon Spin Qubits. *Nature* **2022**, *608*, 682–686, doi:10.1038/s41586-022-04986-6.
52. Hernández-Gómez, S.; Fabbri, N. Quantum Control for Nanoscale Spectroscopy With Diamond Nitrogen-Vacancy Centers: A Short Review. *Front. Phys.* **2021**, *8*, doi:10.3389/fphy.2020.610868.
53. Jin, K.-H.; Jiang, W.; Sethi, G.; Liu, F. Topological Quantum Devices: A Review. *Nanoscale* **2023**, *15*, 12787–12817, doi:10.1039/D3NR01288C.
54. AbuGhanem, M. Superconducting Quantum Computers: Who Is Leading the Future? 2025.
55. Simon, D.S. Solid-State Qubits. In *Introduction to Quantum Science and Technology*; Simon, D.S., Ed.; Springer Nature Switzerland: Cham, 2025; pp. 715–729 ISBN 978-3-031-81315-3.
56. Leveraging Nanomaterials for Measurements at the Quantum Limit - ProQuest Available online: https://www.proquest.com/openview/7fde4142cf741375e4f13c5b8dbede0a/1?pq-origsite=gscholar&cbl=18750&diss=y (accessed on 22 July 2025).
57. Abraheem, S.M.; Ali, M.E.; Abuali, R.M. Emerging Trends in Quantum Sensors: Applications in Defense and Communication. *Middle East Journal of Pure and Applied Sciences (MEJPAS)* **2025**, 19–37.
58. Kaur, T.; Peace, D.; Romero, J. On-Chip High-Dimensional Entangled Photon Sources. *J. Opt.* **2025**, *27*, 023001, doi:10.1088/2040-8986/ada0c5.
59. Butt, M.A.; Imran Akca, B.; Mateos, X. Integrated Photonic Biosensors: Enabling Next-Generation Lab-on-a-Chip Platforms. *Nanomaterials* **2025**, *15*, 731, doi:10.3390/nano15100731.
60. Hosseini, M. Silicon Germanium BiCMOS Integrated Circuits for Scalable Cryogenic Sensing Applications, University of Massachusetts Amherst, 2022.
61. Sadhu, B.; Tien, K.; Chakraborty, S.; Frank, D.; Rosno, P.; Moertl, D.; Yeck, M.; Bulzacchelli, J.; Frolov, D.; Underwood, D.; et al. Cryogenic CMOS Circuits for Future Scaled Quantum Computing Systems: Challenges and Solutions. In Proceedings of the 2025 IEEE Custom Integrated Circuits Conference (CICC); April 2025; pp. 1–3.
62. Brennan, J.C.; Barbosa, J.; Li, C.; Ahmad, M.; Imroze, F.; Rose, C.; Karar, W.; Stanley, M.; Heidari, H.; Ridler, N.M.; et al. Classical Interfaces for Controlling Cryogenic Quantum Computing Technologies 2025.
63. Elshaari, A.W.; Pernice, W.; Srinivasan, K.; Benson, O.; Zwiller, V. Hybrid Integrated Quantum Photonic Circuits. *Nat. Photonics* **2020**, *14*, 285–298, doi:10.1038/s41566-020-0609-x.
64. Kim, J.-H.; Aghaeimeibodi, S.; Carolan, J.; Englund, D.; Waks, E. Hybrid Integration Methods for On-Chip Quantum Photonics. *Optica, OPTICA* **2020**, *7*, 291–308, doi:10.1364/OPTICA.384118.
65. Burkard, G.; Ladd, T.D.; Pan, A.; Nichol, J.M.; Petta, J.R. Semiconductor Spin Qubits. *Rev. Mod. Phys.* **2023**, *95*, 025003, doi:10.1103/RevModPhys.95.025003.
66. Siddiqi, I. Engineering High-Coherence Superconducting Qubits. *Nat Rev Mater* **2021**, *6*, 875–891, doi:10.1038/s41578-021-00370-4.
67. Nanophotonic Integration and Engineering of Defect Qubits in Diamond - ProQuest Available online: https://www.proquest.com/openview/546efb33185488642fc5d08a7d9e12fd/1?pq-origsite=gscholar&cbl=18750&diss=y (accessed on 22 July 2025).
68. Lange, W. Quantum Computing with Trapped Ions. In *Encyclopedia of Complexity and Systems Science*; Springer, New York, NY, 2009; pp. 7218–7249 ISBN 978-0-387-30440-3.
69. Márquez Seco, A.; Takahashi, H.; Keller, M. Novel Ion Trap Design for Strong Ion-Cavity Coupling. *Atoms* **2016**, *4*, 15, doi:10.3390/atoms4020015.
70. Scarlino, P.; van Woerkom, D.J.; Mendes, U.C.; Koski, J.V.; Landig, A.J.; Andersen, C.K.; Gasparinetti, S.; Reichl, C.; Wegscheider, W.; Ensslin, K.; et al. Coherent Microwave-Photon-Mediated Coupling between a Semiconductor and a Superconducting Qubit. *Nat Commun* **2019**, *10*, 3011, doi:10.1038/s41467-019-10798-6.

71. Maurand, R.; Jehl, X.; Kotekar-Patil, D.; Corna, A.; Bohuslavskyi, H.; Laviéville, R.; Hutin, L.; Barraud, S.; Vinet, M.; Sanquer, M.; et al. A CMOS Silicon Spin Qubit. *Nat Commun* **2016**, *7*, 13575, doi:10.1038/ncomms13575.
72. Nitrogen-Vacancy Center. *Wikipedia* 2025.
73. Samad, M.; Shimizu, M.; Hijikata, Y. Demonstration of Quantum Polarized Microscopy Using an Entangled-Photon Source. *Photonics* **2025**, *12*, 127, doi:10.3390/photonics12020127.
74. Degen, C.L.; Reinhard, F.; Cappellaro, P. Quantum Sensing. *Rev. Mod. Phys.* **2017**, *89*, 035002, doi:10.1103/RevModPhys.89.035002.
75. Xavier, J.; Yu, D.; Jones, C.; Zossimova, E.; Vollmer, F. Quantum Nanophotonic and Nanoplasmonic Sensing: Towards Quantum Optical Bioscience Laboratories on Chip. *Nanophotonics* **2021**, *10*, 1387–1435, doi:10.1515/nanoph-2020-0593.
76. Lee, C.; Lawrie, B.; Pooser, R.; Lee, K.-G.; Rockstuhl, C.; Tame, M. Quantum Plasmonic Sensors. *Chem. Rev.* **2021**, *121*, 4743–4804, doi:10.1021/acs.chemrev.0c01028.
77. Molaei, M.J. Principles, Mechanisms, and Application of Carbon Quantum Dots in Sensors: A Review. *Anal. Methods* **2020**, *12*, 1266–1287, doi:10.1039/C9AY02696G.
78. Schirhagl, R.; Chang, K.; Loretz, M.; Degen, C.L. Nitrogen-Vacancy Centers in Diamond: Nanoscale Sensors for Physics and Biology. *Annual Review of Physical Chemistry* **2014**, *65*, 83–105, doi:10.1146/annurev-physchem-040513-103659.
79. Miller, B.S.; Bezinge, L.; Gliddon, H.D.; Huang, D.; Dold, G.; Gray, E.R.; Heaney, J.; Dobson, P.J.; Nastouli, E.; Morton, J.J.L.; et al. Spin-Enhanced Nanodiamond Biosensing for Ultrasensitive Diagnostics. *Nature* **2020**, *587*, 588–593, doi:10.1038/s41586-020-2917-1.
80. Bucher, D.B.; Glenn, D.R.; Park, H.; Lukin, M.D.; Walsworth, R.L. Hyperpolarization-Enhanced NMR Spectroscopy with Femtomole Sensitivity Using Quantum Defects in Diamond. *Phys. Rev. X* **2020**, *10*, 021053, doi:10.1103/PhysRevX.10.021053.
81. Ahmadivand, A.; Gerislioglu, B.; Ramezani, Z.; Kaushik, A.; Manickam, P.; Ghoreishi, S.A. Femtomolar-Level Detection of SARS-CoV-2 Spike Proteins Using Toroidal Plasmonic Metasensors 2020.
82. Hildebrandt, N. Biofunctional Quantum Dots: Controlled Conjugation for Multiplexed Biosensors. *ACS Nano* **2011**, *5*, 5286–5290, doi:10.1021/nn2023123.
83. Zeng, S.; Baillargeat, D.; Ho, H.-P.; Yong, K.-T. Nanomaterials Enhanced Surface Plasmon Resonance for Biological and Chemical Sensing Applications. *Chem. Soc. Rev.* **2014**, *43*, 3426, doi:10.1039/c3cs60479a.
84. Abdel-Salam, M.; Omran, B.; Whitehead, K.; Baek, K.-H. Superior Properties and Biomedical Applications of Microorganism-Derived Fluorescent Quantum Dots. *Molecules* **2020**, *25*, 4486, doi:10.3390/molecules25194486.
85. Pang, X.; Zhang, Y.; Pan, J.; Zhao, Y.; Chen, Y.; Ren, X.; Ma, H.; Wei, Q.; Du, B. A Photoelectrochemical Biosensor for Fibroblast-like Synoviocyte Cell Using Visible Light-Activated NCQDs Sensitized-ZnO/CH3NH3PbI3 Heterojunction. *Biosensors and Bioelectronics* **2016**, *77*, 330–338, doi:10.1016/j.bios.2015.09.047.
86. Roushani, M.; Ghanbari, K.; Jafar Hoseini, S. Designing an Electrochemical Aptasensor Based on Immobilization of the Aptamer onto Nanocomposite for Detection of the Streptomycin Antibiotic. *Microchemical Journal* **2018**, *141*, 96–103, doi:10.1016/j.microc.2018.05.016.
87. Muthusankar, G.; Devi, R.K.; Gopu, G. Nitrogen-Doped Carbon Quantum Dots Embedded Co3O4 with Multiwall Carbon Nanotubes: An Efficient Probe for the Simultaneous Determination of Anticancer and Antibiotic Drugs. *Biosensors and Bioelectronics* **2020**, *150*, 111947, doi:10.1016/j.bios.2019.111947.
88. Saadati, A.; Hassanpour, S.; Bahavarnia, F.; Hasanzadeh, M. A Novel Biosensor for the Monitoring of Ovarian Cancer Tumor Protein CA 125 in Untreated Human Plasma Samples Using a Novel Nano-Ink: A New Platform for Efficient Diagnosis of Cancer Using Paper Based Microfluidic Technology. *Anal. Methods* **2020**, *12*, 1639–1649, doi:10.1039/D0AY00299B.
89. Serafín, V.; Valverde, A.; Garranzo-Asensio, M.; Barderas, R.; Campuzano, S.; Yáñez-Sedeño, P.; Pingarrón, J.M. Simultaneous Amperometric Immunosensing of the Metastasis-Related Biomarkers IL-13Rα2 and CDH-17 by Using Grafted Screen-Printed Electrodes and a Composite Prepared from Quantum Dots and Carbon Nanotubes for Signal Amplification. *Microchim Acta* **2019**, *186*, 411, doi:10.1007/s00604-019-3531-5.
90. Sánchez Toural, J.L.; Marzoa, V.; Bernardo-Gavito, R.; Pau, J.L.; Granados, D. Hands-On Quantum Sensing with NV− Centers in Diamonds. *C* **2023**, *9*, 16, doi:10.3390/c9010016.

91. Sharmin, R.; Hamoh, T.; Sigaeva, A.; Mzyk, A.; Damle, V.G.; Morita, A.; Vedelaar, T.; Schirhagl, R. Fluorescent Nanodiamonds for Detecting Free-Radical Generation in Real Time during Shear Stress in Human Umbilical Vein Endothelial Cells. *ACS Sens.* **2021**, *6*, 4349–4359, doi:10.1021/acssensors.1c01582.

92. Li, C.; Soleyman, R.; Kohandel, M.; Cappellaro, P. SARS-CoV-2 Quantum Sensor Based on Nitrogen-Vacancy Centers in Diamond. *Nano Lett.* **2022**, *22*, 43–49, doi:10.1021/acs.nanolett.1c02868.

93. B. Baraeinejad et al., "Clinical IoT in Practice: A Novel Design and Implementation of a Multi-functional Digital Stethoscope for Remote Health Monitoring," TechRxiv Preprints, Nov. 7, 2023, doi: 10.36227/techrxiv.24459988.v1.

94. Baraeinejad, B., Forouzesh, M., et al. (2024). Design and Implementation of an IoT-based Respiratory Motion Sensor. arXiv preprint arXiv:2412.05405.

95. Y. Torabi et al., "Exploring Sensing Devices for Heart and Lung Sound Monitoring," arXiv preprint, doi: 2406.12432, 2024.

96. Torabi, Y.; Shirani, S.; Reilly, J.P.; Gauvreau, G.M. MEMS and ECM Sensor Technologies for Cardiorespiratory Sound Monitoring—A Comprehensive Review. *Sensors (Basel)* **2024**, *24*, 7036, doi:10.3390/s24217036.

97. Jang, B.; Cao, P.; Chevalier, A.; Ellington, A.; Hassibi, A. A CMOS Fluorescent-Based Biosensor Microarray. In Proceedings of the 2009 IEEE International Solid-State Circuits Conference - Digest of Technical Papers; February 2009; pp. 436-437,437a.

98. Li, H.; Parsnejad, S.; Ashoori, E.; Thompson, C.; Purcell, E.K.; Mason, A.J. Ultracompact Microwatt CMOS Current Readout With Picoampere Noise and Kilohertz Bandwidth for Biosensor Arrays. *IEEE Transactions on Biomedical Circuits and Systems* **2018**, *12*, 35–46, doi:10.1109/TBCAS.2017.2752742.

99. Mehdipoor, M.; Ghavifekr, H.B. A Novel Microfluidics Integrated Biosensor Based on a MEMS Resonator. *Microsyst Technol* **2020**, *26*, 3821–3828, doi:10.1007/s00542-020-04870-1.

100. Sang, S.; Witte, H. A Novel PDMS Micro Membrane Biosensor Based on the Analysis of Surface Stress. *Biosensors and Bioelectronics* **2010**, *25*, 2420–2424, doi:10.1016/j.bios.2010.03.035.

101. Kurmendra; Kumar, R. MEMS Based Cantilever Biosensors for Cancer Detection Using Potential Bio-Markers Present in VOCs: A Survey. *Microsyst Technol* **2019**, *25*, 3253–3267, doi:10.1007/s00542-019-04326-1.

102. Timurdogan, E.; Alaca, B.E.; Kavakli, I.H.; Urey, H. MEMS Biosensor for Detection of Hepatitis A and C Viruses in Serum. *Biosensors and Bioelectronics* **2011**, *28*, 189–194, doi:10.1016/j.bios.2011.07.014.

103. Bharati, M.; Rana, L.; Gupta, R.; Sharma, A.; Jha, P.K.; Tomar, M. Realization of a DNA Biosensor Using Inverted Lamb Wave MEMS Resonator Based on ZnO/SiO2/Si/ZnO Membrane. *Analytica Chimica Acta* **2023**, *1249*, 340929, doi:10.1016/j.aca.2023.340929.

104. Silicon Photonics Circuit Design: Methods, Tools and Challenges - Bogaerts - 2018 - Laser & Photonics Reviews - Wiley Online Library Available online: https://onlinelibrary.wiley.com/doi/full/10.1002/lpor.201700237?casa_token=-QraakcNN34AAAAA%3AATwiaMvGhO3eUQBB7Qvq2Mie-PHDmWnDgbr5f6qctFqmftYZ7s-ZJVSVX4EaG-if79n7PLXe78JlQuzE (accessed on 23 July 2025).

105. Bogaerts, W.; Baets, R.; Dumon, P.; Wiaux, V.; Beckx, S.; Taillaert, D.; Luyssaert, B.; Van Campenhout, J.; Bienstman, P.; Van Thourhout, D. Nanophotonic Waveguides in Silicon-on-Insulator Fabricated with CMOS Technology. *Journal of Lightwave Technology* **2005**, *23*, 401–412, doi:10.1109/JLT.2004.834471.

106. Juan Colás, J. Introduction to Label-Free Biosensing. In *Dual-Mode Electro-photonic Silicon Biosensors*; Juan Colás, J., Ed.; Springer International Publishing: Cham, 2017; pp. 7–35 ISBN 978-3-319-60501-2.

107. Claes, T. Advanced Silicon Photonic Ring Resonator Label-Free Biosensors. dissertation, Ghent University, 2012.

108. Heideman, R.G.; Lambeck, P.V. Remote Opto-Chemical Sensing with Extreme Sensitivity: Design, Fabrication and Performance of a Pigtailed Integrated Optical Phase-Modulated Mach–Zehnder Interferometer System. *Sensors and Actuators B: Chemical* **1999**, *61*, 100–127, doi:10.1016/s0925-4005(99)00283-x.

109. Vogelbacher, F.; Kothe, T.; Muellner, P.; Melnik, E.; Sagmeister, M.; Kraft, J.; Hainberger, R. Waveguide Mach-Zehnder Biosensor with Laser Diode Pumped Integrated Single-Mode Silicon Nitride Organic Hybrid Solid-State Laser. *Biosensors and Bioelectronics* **2022**, *197*, 113816, doi:10.1016/j.bios.2021.113816.

110. Densmore, A.; Vachon, M.; Xu, D.-X.; Janz, S.; Ma, R.; Li, Y.-H.; Lopinski, G.; Delâge, A.; Lapointe, J.; Luebbert, C.C.; et al. Silicon Photonic Wire Biosensor Array for Multiplexed Real-Time and Label-Free Molecular Detection. *Opt. Lett.* **2009**, *34*, 3598, doi:10.1364/ol.34.003598.

111. Crespi, A.; Gu, Y.; Ngamsom, B.; Hoekstra, H.J.W.M.; Dongre, C.; Pollnau, M.; Ramponi, R.; Vlekkert, H.H. van den; Watts, P.; Cerullo, G.; et al. Three-Dimensional Mach-Zehnder Interferometer in a Microfluidic Chip for Spatially-Resolved Label-Free Detection. *Lab Chip* **2010**, *10*, 1167–1173, doi:10.1039/B920062B.

112. Kumar, S.; Kumar, A.; Mishra, R.D.; Babu, P.; Pandey, S.K.; Pal, M.K.; Kumar, M. Nanophotonic Ring Resonator Based on Slotted Hybrid Plasmonic Waveguide for Biochemical Sensing. *IEEE Sensors Journal* **2023**, *23*, 5695–5702, doi:10.1109/JSEN.2023.3239868.

113. Singh, R.; Chack, D.; Priye, V. SNROW-Based Highly Sensitive Label-Free Surface Biosensor for Hepatitis B Detection. *Appl. Opt., AO* **2022**, *61*, 6510–6517, doi:10.1364/AO.463800.

114. Haron, M.H.; Berhanuddin, D.D.; Shaari, S.; Yeop Majlis, B.; Md Zain, A.R. The Design of Tunable Photonic Crystal Biosensor With the Integration of PN Phase Shifter Using PIC Design Approach 2021.

115. Voronkov, G.; Zakoyan, A.; Ivanov, V.; Iraev, D.; Stepanov, I.; Yuldashev, R.; Grakhova, E.; Lyubopytov, V.; Morozov, O.; Kutluyarov, R. Design and Modeling of a Fully Integrated Microring-Based Photonic Sensing System for Liquid Refractometry. *Sensors* **2022**, *22*, 9553, doi:10.3390/s22239553.

116. Carroll, L.; Lee, J.-S.; Scarcella, C.; Gradkowski, K.; Duperron, M.; Lu, H.; Zhao, Y.; Eason, C.; Morrissey, P.; Rensing, M.; et al. Photonic Packaging: Transforming Silicon Photonic Integrated Circuits into Photonic Devices. *Applied Sciences* **2016**, *6*, 426, doi:10.3390/app6120426.

117. Yang, J.; Tang, M.; Chen, S.; Liu, H. From Past to Future: On-Chip Laser Sources for Photonic Integrated Circuits. *Light Sci Appl* **2023**, *12*, 16, doi:10.1038/s41377-022-01006-0.

118. Lu, L.; Zheng, X.; Lu, Y.; Zhu, S.; Ma, X.-S. Advances in Chip-Scale Quantum Photonic Technologies. *Advanced Quantum Technologies* **2021**, *4*, 2100068, doi:10.1002/qute.202100068.

119. Li, R.; Liu, F.; Lu, Q. Quantum Light Source Based on Semiconductor Quantum Dots: A Review. *Photonics* **2023**, *10*, 639, doi:10.3390/photonics10060639.

120. Silverstone, J.W.; Bonneau, D.; Ohira, K.; Suzuki, N.; Yoshida, H.; Iizuka, N.; Ezaki, M.; Natarajan, C.M.; Tanner, M.G.; Hadfield, R.H.; et al. On-Chip Quantum Interference between Silicon Photon-Pair Sources. *Nature Photon* **2014**, *8*, 104–108, doi:10.1038/nphoton.2013.339.

121. Ezawa, M. Electric Circuits for Universal Quantum Gates and Quantum Fourier Transformation. *Phys. Rev. Res.* **2020**, *2*, 023278, doi:10.1103/PhysRevResearch.2.023278.

122. Grottke, T.; Hartmann, W.; Schuck, C.; Pernice, W.H.P. Optoelectromechanical Phase Shifter with Low Insertion Loss and a 13π Tuning Range. *Opt. Express, OE* **2021**, *29*, 5525–5537, doi:10.1364/OE.413202.

123. Li, Z.; Jin, X.; Yuan, C.; Wang, K. Photon Detector Technology for Laser Ranging: A Review of Recent Developments. *Coatings* **2025**, *15*, 798, doi:10.3390/coatings15070798.

124. Dao, T.H.; Amanti, F.; Andrini, G.; Armani, F.; Barbato, F.; Bellani, V.; Bonaiuto, V.; Cammarata, S.; Campostrini, M.; Cornia, S.; et al. Single-Photon Detectors for Quantum Integrated Photonics. *Photonics* **2025**, *12*, 8, doi:10.3390/photonics12010008.

125. Na, N.; Lu, Y.-C.; Liu, Y.-H.; Chen, P.-W.; Lai, Y.-C.; Lin, Y.-R.; Lin, C.-C.; Shia, T.; Cheng, C.-H.; Chen, S.-L. Room Temperature Operation of Germanium–Silicon Single-Photon Avalanche Diode. *Nature* **2024**, *627*, 295–300, doi:10.1038/s41586-024-07076-x.

126. Holzman, I.; Ivry, Y. Superconducting Nanowires for Single-Photon Detection: Progress, Challenges, and Opportunities. *Advanced Quantum Technologies* **2019**, *2*, 1800058, doi:10.1002/qute.201800058.

127. Hao, H.; Zhao, Q.-Y.; Huang, Y.-H.; Deng, J.; Yang, F.; Ru, S.-Y.; Liu, Z.; Wan, C.; Liu, H.; Li, Z.-J.; et al. A Compact Multi-Pixel Superconducting Nanowire Single-Photon Detector Array Supporting Gigabit Space-to-Ground Communications. *Light Sci Appl* **2024**, *13*, 25, doi:10.1038/s41377-023-01374-1.

128. Stosch, J.H.; Kühler, T.; Griese, E. Optical Directional Coupler for Graded Index Waveguides in Thin Glass Sheets for PCB Integration. In Proceedings of the 2016 IEEE 20th Workshop on Signal and Power Integrity (SPI); May 2016; pp. 1–4.

129. Righini, G.C.; Liñares, J. Active and Quantum Integrated Photonic Elements by Ion Exchange in Glass. *Applied Sciences* **2021**, *11*, 5222, doi:10.3390/app11115222.